\documentclass[fleqn,usenatbib]{mnras}

\usepackage{newtxtext,newtxmath}

\usepackage[T1]{fontenc}

\DeclareRobustCommand{\VAN}[3]{#2}
\let\VANthebibliography\thebibliography
\def\thebibliography{\DeclareRobustCommand{\VAN}[3]{##3}\VANthebibliography}

\usepackage{graphicx}	%
\usepackage{amsmath}	%
\usepackage{siunitx}
\usepackage{footnote}
\usepackage{multirow}

\newcommand{\lya}{Ly$\alpha$}
\newcommand{\lyaf}{Ly$\alpha$ forest}
\newcommand{\mpch}{\ensuremath{h^{-1}\,\mathrm{cMpc}}}

\newcommand{\mpc}{\ensuremath{\mathrm{cMpc}}}

\newcommand{\msun}{\ensuremath{M_\odot}}
\newcommand{\msunh}{\ensuremath{h^{-1}\,M_\odot}}

\newcommand{\reducedchisq}{\ensuremath{\chi^2_r}}
\newcommand{\blya}{\ensuremath{b_{\text{\lya{}}}}}

\newcommand{\angstrom}{\text{\normalfont\AA}}
\newcommand{\kms}{\text{\normalfont{km/s}}}

\newcommand{\threedhst}{3D-HST}
\newcommand{\clamato}{CLAMATO}
\newcommand{\mosdef}{MOSDEF}
\newcommand{\vuds}{VUDS}
\newcommand{\zdeep}{zCOSMOS-deep}

\newcommand{\vega}{\texttt{Vega}}

\newcommand{\bgal}{\ensuremath{b_{\mathrm{g}}}}
\newcommand{\deltaz}{\ensuremath{\delta_z}}
\newcommand{\sigz}{\ensuremath{\sigma_z}}
\newcommand{\lgmstar}{\ensuremath{\log_{10}(M_* / \msun{})}}

\newcommand{\alld}{3D}
\newcommand{\projd}{2D}

\title[COSMOS LyA forest-galaxy cross-correlations]{Cross-correlations between the CLAMATO Lyman-alpha forest and galaxies within the COSMOS field}

\author[Zhang et al.]{
Benjamin Zhang$^{1}$,
Khee-Gan Lee$^{2,3}$\thanks{kglee@ipmu.jp},
Andrei Cuceu$^{4,5,6}$,
Andreu Font-Ribera$^{7,8}$,
Rieko Momose$^{9,10}$
\\
$^{1}$Department of Physics and Astronomy, University of Southern California, Los Angeles, CA 90089, USA\\
$^{2}$Kavli IPMU (WPI), UTIAS, The University of Tokyo, Kashiwa, Chiba 277-8583, Japan\\
$^{3}$Center for Data-Driven Discovery, Kavli IPMU (WPI), UTIAS, The University of Tokyo, Kashiwa, Chiba 277-8583, Japan \\
$^{4}$Lawrence Berkeley National Laboratory, 1 Cyclotron Road, Berkeley, CA 94720, USA\\
$^{5}$Center for Cosmology and AstroParticle Physics, The Ohio State University, 191 West Woodruff Avenue, Columbus, OH 43210, USA\\
$^{6}$NASA Einstein Fellow\\
$^{7}$Institut de Física d’Altes Energies (IFAE), The Barcelona Institute of Science and Technology, Edifici Cn, Campus UAB, 08193, Bellaterra (Barcelona), Spain\\
$^{8}$Institució Catalana de Recerca i Estudis Avançats, Passeig de Lluís Companys, 23, 08010, Barcelona, Spain\\
$^{9}$Carnegie Observatories, 813 Santa Barbara Street, Pasadena, CA 91101, USA\\
$^{10}$Department of Astronomy, School of Science, The University of Tokyo, 7-3-1 Hongo, Bunkyo-ku, Tokyo, 113-0033, Japan\\
}

\date{Accepted XXX. Received YYY; in original form ZZZ}

\pubyear{\the\year{}}

\begin{document}
\label{firstpage}
\pagerange{\pageref{firstpage}--\pageref{lastpage}}
\maketitle

\begin{abstract}
    We compute the 3D cross-correlation between the absorption of the $z\sim 2.3$ Lyman-alpha forest measured by the COSMOS Lyman-Alpha Mapping And Tomography Observations (CLAMATO) survey, and 1642 foreground galaxies with spectroscopic redshifts from several different surveys, including 3D-HST, CLAMATO, zCOSMOS-deep, MOSDEF, and VUDS. 
    For each survey, we compare the measured cross-correlation with models incorporating the galaxy linear bias as well as observed redshift dispersion and systematic redshift offset. 
    The derived redshift dispersion and offsets are generally consistent with those expected from, e.g., spectroscopic redshifts measured with UV absorption lines or NIR emission lines observed with specific instruments, but we find hints of `fingers-of-god' caused by overdensities in the field. 
    We combine our foreground galaxy sample, and split them into 3 bins of robustly-estimated stellar mass in order to study the stellar mass-halo mass relationship. 
    For sub-samples with median stellar masses of $\log_{10}(M_* / M_\odot) = [9.28,9.74,10.22]$, we find galaxy biases of $b_g\approx [2.1, 3.2,3.8]$, respectively.
    A comparison with mock measurements from the Bolshoi-Planck $N$-body simulation yields corresponding halo masses of $\log_{10}(M_* / M_\odot) \approx [10.5,11.7,12.1]$ for these stellar mass bins.
    At the low mass end, our results suggest enhanced star formation histories in mild tension with predictions from previous angular correlation and abundance matching-based observations, and the IllustrisTNG simulation.
\end{abstract}


\section{Introduction}
\label{sec:intro}

The hydrogen Lyman-$\alpha$ (Ly$\alpha$) forest, characterized by numerous absorption features observed in the spectra of distant quasars and galaxies, has proven to be a crucial probe of large-scale cosmic structure. These absorption lines arise primarily from intervening diffuse neutral hydrogen gas in the intergalactic medium (IGM), which trace density and ionization fluctuations on scales ranging from sub-Mpc to tens of Mpc \citep{gunn_1965_lyaforest, lynds_1971_lyaforest, rauch_1998_lyaforest, croft_2002_lyaforest}. At redshifts $z \sim 2-3$, the Ly$\alpha$ forest traces the matter density at the epoch of ``cosmic noon'', which corresponds to the peak of star formation and quasar activity.
Thereby, it provides valuable constraints on cosmological parameters (for example, through measurements of baryon acoustic oscillations), as well the underlying matter power spectrum \citep{hernquist_1996_cdmlya, viel_2004_lyapowerspec, slosar_2011_lyaforest_boss}.

Cross-correlations between the Ly$\alpha$ forest and other cosmological tracers, such as galaxies, quasars, or damped Ly$\alpha$ systems (DLAs), have emerged as powerful tools for investigating the connection between the diffuse IGM and collapsed structures. 
\citet{fontribera-2013-lya-quasar-xcorr, fontribera-2014-boss-xcorr} demonstrated that cross-correlations between the Ly$\alpha$ forest and quasars provide constraints on quasar clustering properties and cosmological parameters, while simultaneously allowing investigations into the quasar-hosting halos' environments.
Furthermore, cross-correlation constraints on cosmological parameters like the BAO scale are complementary to constraints from \lya{} autocorrelation measurements, yielding better constraints when combined \citep{fontribera-2014-boss-xcorr, blomqvist-2019-ebossdr14-crosscorr, bourboux-20-eboss}.

Traditionally, studies of the Ly$\alpha$ forest have utilized quasars as background sources owing to their brightness and distinctive spectral features. However, observational efforts over the past decade have enabled the use of galaxies, particularly star-forming Lyman-break galaxies (LBGs), as alternative background sources. The COSMOS Lyman-Alpha Mapping And Tomography Observations (CLAMATO) survey has pioneered this approach by utilizing dense spectroscopic sampling of LBGs at $z \sim 2-3$ to reconstruct the three-dimensional distribution of the Ly$\alpha$ forest with unprecedented spatial resolution \citep{lee-2014-lya-tomography, lee-2018-clamato-dr1, horowitz-22-clamato-dr2}. The CLAMATO data has allowed detailed mapping of the IGM on scales down to several Mpc, revealing structures such as galaxy protoclusters \citep{lee-2016-clamato-dr1-protocluster, newman-2022-protoclusters}, cosmic voids \citep{Krolewski-2018-clamato-voids}, and the filamentary cosmic web \citep{horowitz-19-tardis-1,horowitz-21-tardis-2,horowitz-22-clamato-dr2}.
With LBGs as background sources, the statistical power of cross-correlations is significantly improved on small scales $\sim 1\,$Mpc, and the targeted fields are also likely to be within smaller, well-observed regions with significant multi-wavelength data. 
This potentially opens up additional information on the cross-correlated objects such as their stellar masses, star-formation rates, and AGN activity. 

\citet{momose:2021} had --- using an earlier data release of CLAMATO data --- carried out a cross-correlation between the \lya{} absorption with galaxy properties in the COSMOS field. 
However, they had done so by stacking the \lya{} flux values from the Wiener-filtered map at the galaxy. This makes it more challenging to fit models to the observations due to the Wiener-filtering step, which imposes a discretization due to the gridding of the flux field, among other systematic effects. 

Recently, \citet{newman-24-latis} published a cross-correlation analysis between their LATIS \lya{} survey data of $\sim3000$ background sightlines, with the foreground LBG redshifts seen within the same survey. %
Using the ASTRID hydrodynamic simulation to connect the observed cross-correlation profiles of different galaxy stellar mass bins with the bins' halo masses, they obtain constraints on the stellar halo-mass relation (SMHR) at $z \sim 2.5$.
They report constraints that are consistent with those obtained from galaxy-galaxy autocorrelation in the same work, as well as with simulations and photometric-redshift observations.

In this paper, we carry out a cross-correlation study of galaxies residing near the CLAMATO \lya{} absorption data in the COSMOS field. This represents an alternative analysis in a similar vein as \citet{newman-24-latis}, but also differs in several ways. Firstly, in addition to the foreground LBGs observed within the CLAMATO survey itself, 
we also exploit the treasury of spectroscopic redshifts that have been accumulated within the COSMOS field, 
which include observations from various instruments ranging traditional optical slit-spectrographs, near-infrared (NIR) spectroscopy, and \textit{Hubble Space Telescope} (HST) grism spectroscopy. 
Secondly, our analysis first obtains constraints on the linear galaxy bias, as well as the systemic spectroscopic redshift offsets and scatter of the various galaxy surveys.
We then convert these into constraints on the halo-stellar mass relation. This is in contrast with \citet{newman-24-latis}, which directly fits for the halo-stellar mass relationship from the observed cross-correlation using numerical simulations.
Finally, our analysis combines data from several galaxy surveys with differing spectroscopic redshift accuracies, which presents unique challenges as the fitted redshift accuracy can be degenerate with the linear galaxy bias.

    Unless specifically stated otherwise, throughout this work we adopt the Planck 2015 $\Lambda$CDM concordance cosmology \citep{planck_2015}.

\section{Data}
\label{sec:data}
In this paper, we analyze data covering a footprint of $0.2\,\mathrm{deg}^2$ within the COSMOS field, anchored on the $z\approx 2.0-2.5$ Lyman-$\alpha$ forest data obtained by the COSMOS Lyman-Alpha Mapping And Tomography Observations (CLAMATO) survey, cross-correlated with spectroscopic galaxy redshift samples obtained by various surveys within the COSMOS field \citep{scoville2007,capak2007}.

    \subsection{Lyman-$\alpha$ forest data}
    \label{sec:data:lya}
    The \lya{} forest data in this paper are from the 2nd (and final) data release of the CLAMATO Survey \citep{horowitz-22-clamato-dr2} which were observed using the LRIS spectrograph on the Keck-I telescope at the W.M. Keck Observatory on Maunakea, Hawai`i. This was the first survey specifically designed to target star-forming UV-bright Lyman-Break Galaxies (LBGs) as background sources probing the foreground \lya{} forest \citep{lee2014,lee2014a,lee-2018-clamato-dr1}, instead of the comparatively brighter quasars that have traditionally been used. 

    We use 340 galaxy and quasar spectra from CLAMATO that cover a footprint of approximately 0.2 deg$^2$ in the center of the COSMOS field; these include sightlines nominally outside the footprint of the tomographic map presented in \citet{horowitz-22-clamato-dr2}. The spectra have a typical spectral resolution of $R\equiv \lambda/\Delta\lambda \sim 1100$ and a signal-to-noise selected to be at least S/N$\sim 2$ per pixel within their restframe $1040\angstrom\leq \lambda_\mathrm{rest} \leq 1195\angstrom$ Ly$\alpha$ forest region.
    The foreground Lyman-$\alpha$ forest in these sources have at least some coverage at $2.05\leq z_\alpha < 2.55$ (the redshift extent of the \citealt{horowitz-22-clamato-dr2} maps), but in the cross-correlation analysis we will use all Ly$\alpha$ pixels in the background sources including those that lie below or above the aforementioned range.
  
    On the reduced 1D spectra, the mean-flux regulation \citep{lee2012} technique was carried out to estimate the intrinsic continuum level $C$.
    Given the observed spectral flux density, $f$, of the spectrum, the Ly$\alpha$ forest fluctuation was then derived for each pixel:
    \begin{equation}
        \delta_F =\frac{f}{C\, \langle F \rangle (z)} -1, 
    \end{equation}
    where $\langle F \rangle (z)$ is the mean \lyaf{} transmission at the redshift $z$.
    The \lyaf{} data for each sightline is saved as vectors listing the redshift and $\delta_F$, as well the R.A.\ and Decl.\ coordinates on the celestial plane.

    There is one significant difference in how we treat the data here, compared with the past CLAMATO publications. 
    When defining the 3D comoving grid for the IGM tomographic reconstructions, the past analyses had assumed a fixed Hubble parameter, $H(z=2.30)$, such that the differential comoving distance $\mathrm{d}\chi/\mathrm{d}z$ is static across the entire $2.05\leq z_\alpha < 2.55$  of the map volume.
    This defined a one-to-one correspondence between the Cartesian $x$ and $y$ axes with the Real Ascension and Declination at each sightline, 
    but at the expense of several-percent errors at the lower and upper ends of the mapped redshift range.
    In other words, the usual flared geometry that one expects from a lightcone with fixed angular footprint had been flattened into a rectangular shape.
    For the present cross-correlation analysis there is no need to make the approximation of fixed $H(z)$, and so we always compute the correct transverse and line-of-sight comoving distances between each Ly$\alpha$ forest pixel and galaxy, assuming our fiducial cosmology.
    
    \subsection{Galaxy redshift survey data}
    \label{sec:data:gal}

    We cross-correlate the the CLAMATO \lyaf{} data with a sample of foreground galaxies drawn not just from CLAMATO itself, but including various other spectroscopic surveys that have targeted the COSMOS field. 

    The CLAMATO \lyaf{} was primarily intended to create a tomographic map at $2.05 \leq z_\alpha < 2.55$, but some sightlines probe slightly beyond these boundaries. In addition, our cross-correlation analysis probes scales of up to $\sim 30\,\mathrm{cMpc}$ in both the transverse and radial dimensions. We therefore select foreground galaxies at $2.0\leq z < 2.6$, beyond the formal CLAMATO redshift range, and allowed them to lie up to $15\,h^{-1}\,\mathrm{Mpc}$ outside the CLAMATO survey footprint on the transverse (i.e. sky) plane since they would still contribute to the large-scale cross-correlation signal. 

In total, we have a total of 1950 galaxies from all the surveys (1642 unique galaxies), used in our cross-correlation analysis. Objects known to be AGN (typically through their broad emission lines and blue continua) were removed from our sample. 

For combined analyses with CLAMATO, we compiled the combined catalog through private communication with the zCOSMOS and VUDS teams, as well as with publicly-available redshifts, with the stellar masses obtained through cross-matching with the COSMOS2020 photometric catalog as described in \citet{momose:2024}. Most of the data in our sample were nominally included in the recent release of COSMOS spectroscopic redshifts by \citet{khostovan-2026-cosmos-compilation}. We have not, however, checked whether our sample is consistent with their compilation.

    \begin{figure*}
        \centering
        \includegraphics[width=0.6\linewidth]{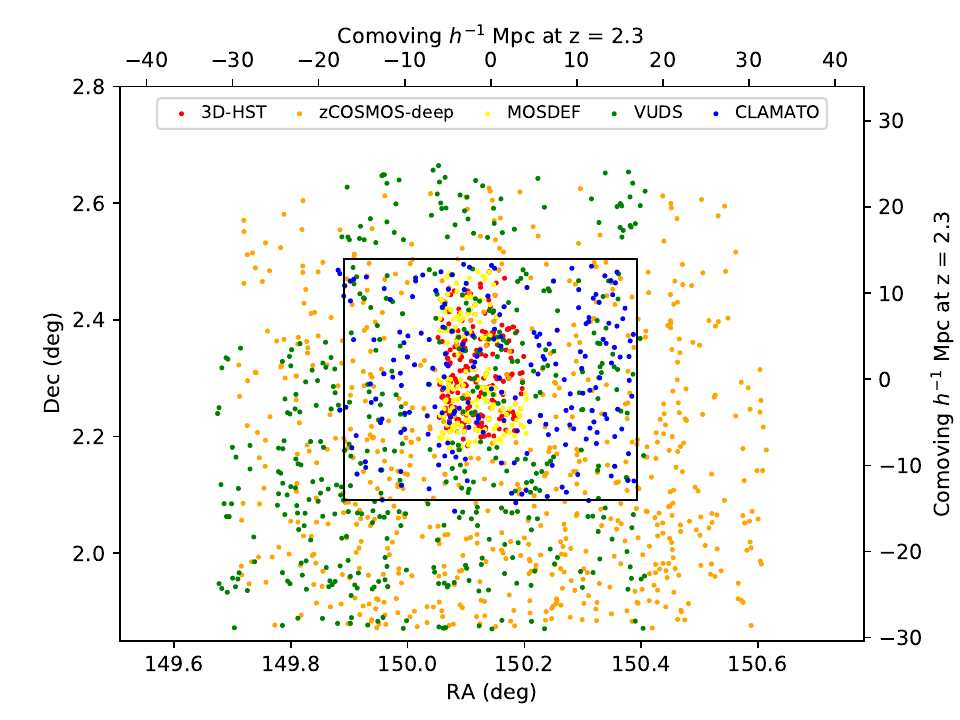}
        \caption{The on-sky distribution of foreground galaxy redshifts used in this work, shown in comparison with the CLAMATO DR2 \lyaf{} survey footprint (black outline). Assuming a midpoint redshift of $z = 2.3$, the conversion to \mpch{} is also shown for reference on the alternate axes, centered on the CLAMATO footprint.}
        \label{fig:gal-spatial-map}
    \end{figure*}

    Here, we briefly describe the different galaxy redshift surveys used, and show their positions on the sky in Figure \ref{fig:gal-spatial-map}.

    \subsubsection*{CLAMATO} 
    The CLAMATO spectroscopic catalog consists of 393 sources with high-confidence redshifts from visual inspection, of which 377 were at $z>2$.
    Of these, 359 were LBGs in the sense that they are relatively UV-bright galaxies that show clear continuum detections. 
    Their selections were partially based on re-targetings of spectroscopically-confirmed sources as well as multi-wavelength photometric redshifts from \citet{ilbert2009}; \citet{laigle2016}; \citet{davidzon2017}.
    The overall magnitude limit was set rather aggressively to $g_\mathrm{AB} < 25.3$ in order to maximize the number of slits in the multi-object slitmasks; however, the majority of the galaxies with successful redshifts were at $g_\mathrm{AB} \leq 24.8$, having received typical integration times of $\approx 2$hrs.
    The CLAMATO systemic redshifts are based on the restframe-UV absorption lines if the continuum signal-to-noise permits, but in some lower-S/N objects they were estimated off the Ly$\alpha$ emission line.
    This mix of estimates from Ly$\alpha$ emission and UV absorption lines introduces some confusion to our estimates of systemic redshift offsets. This is unfortunate, but does not affect the Ly$\alpha$ forest absorption derived from these spectra.
    The spectral resolution of the data is $R \approx 1100$.
    We select galaxies with a redshift confidence flag of 3 or 4, indicating a reasonable- or high-confidence redshift.
    In the cross-correlation analysis, 205 CLAMATO galaxies were used as part of the foreground galaxy sample.
    \subsubsection*{zCOSMOS-deep} 
    zCOSMOS-deep was the deeper component of the overall spectroscopic zCOSMOS survey \citep{lilly_2007_zcosmos} conducted on the VLT-VIMOS optical spectrograph \citep{le-fevre:2003}. The galaxies were selected through several color-color criteria (including `\textit{UGR}' and `\textit{BzK}'; \citealt{Steidel2004} and \citealt{Daddi2004}, respectively), covering the central 1 deg$^2$ footprint of the COSMOS field. 
    There was also an overall magnitude limit of $22.5 < B_\mathrm{AB} < 25.0$.
    The targets were observed for $\sim$ 4-5hrs with the low-resolution ($R\sim 200$) LR-BLUE grism covering a wavelength range of $3600 \angstrom \lesssim \lambda \lesssim 6800\angstrom$.
    Therefore, the zCOSMOS-deep galaxies at $2.0<z<2.6$ are also classed as LBGs identified through both UV absorption lines and Ly$\alpha$ emission, albeit observed at lower resolution than CLAMATO.  
    We selected zCOSMOS-deep galaxies with visual inspection flags of $\geq 2$, of which 749 were included in our cross-correlation.
    \subsubsection{VUDS}
    The VIMOS Ultra-Deep Survey (VUDS; \citealt{lefevre_2015_vuds},\citealt{tasca2016}) was another spectroscopic optical survey carried out with VLT-VIMOS, which covered $\sim$10,000 $2<z<6$ galaxies across multiple fields including 0.8 deg$^2$ in COSMOS. 
    While using the LR-BLUE grism as zCOSMOS-deep (and hence the same spectral resolution, $R\sim 200$), VUDS additionally observed with the LR-RED grism to go up to $\lambda = 9350\,\angstrom$ --- which is of little consequence to $z<3$ LBGs. 
    The most significant difference with zCOSMOS-deep is that VUDS galaxies were observed for a longer integration time of 14hrs, leading to better S/N that enable secure redshifts on fainter galaxies that also generally have lower stellar masses. 
    In addition, the VUDS targets was selected with photometric redshifts rather than color-color selection, which allowed greater selection efficiency at the cost of having a radial selection function that is harder to characterize.
    We select galaxies with a redshift reliability flag of 2-9, indicating a $>75\%$ chance to be correct.
    Our cross-correlation analysis includes 469 galaxies from VUDS.
    \subsubsection{MOSDEF}
    The MOSFIRE Deep Evolution Field (MOSDEF; \citealt{kriek_2015_mosdef}) was a spectroscopic survey carried out on the near-infrared (NIR) using the MOSFIRE NIR spectrograph on the Keck-I telescope. 
    This survey targeted $H$-band ($\bar{\lambda} \sim 1.6\,\mu\mathrm{m}$)-selected galaxies within the CANDELS imaging survey fields, which includes $\sim 144$ arcmin$^2$ in COSMOS wholly covered by the CLAMATO footprint.
    Several redshift ranges were targeted in MOSDEF, of which the $2.09 \leq z \leq 2.61$ range overlaps with CLAMATO --- the magnitude limit for these targets were $H<24.5$. 
    The observations were carried across the $Y, J, H\,\&\, K$ NIR spectral windows with spectral resolutions of $R\sim 3000-3650$, covering the restframe optical wavelengths of the galaxies.
    The vast majority of the MOSDEF redshifts were based on restframe-optical nebular emission lines including [OII], H$\beta$, [OIII], H$\alpha$, [NII], and [SII].
    We select galaxies with redshift quality flag above 2, with 185 galaxies having the highest redshift quality of 7.
    We incorporate 195 MOSDEF galaxies derived from their final (January 2021) data release\footnote{\url{https://mosdef.astro.berkeley.edu/for-scientists/data-releases/}}.

    \subsubsection*{3D-HST}
    3D-HST \citep{brammer_2012_3dhst,momcheva2016} was a NIR grism spectroscopic survey carried out using the WFC3 instrument on board the Hubble Space Telescope. 
    Using the G141 grism covering a wavelength range of $1.1\,\mu\mathrm{m} \leq \lambda \leq 1.65\,\mu \mathrm{m}$, this program targeted $\sim 600$ arcmin$^2$ of area over several of the CANDELS fields, including 122 arcmin$^2$ within COSMOS wholly within the CLAMATO footprint. 
    As a slitless spectroscopic program, no explicit target selection was carried out and a large number of galaxies ($\sim 250,000$) were observed, albeit at a low resolution of $R\sim 100$.
    Due the low resolution, relatively limited wavelength coverage, and the possibility of source blending in overlapping spectra, typically only single emission lines or unresolved emission line doublets/complexes (particularly the [OII]$\lambda$ 3737 and [OIII]$\lambda\lambda$4959+5007 at our redshift of interest) are seen in the grism data, therefore combinations of multi-wavelength photometric data together with the grism information provided the most robust redshifts.
    We use the v4.1.5 grism catalog from COSMOS, downloaded from STScI MAST website\footnote{\url{https://archive.stsci.edu/prepds/3d-hst/}}. 
    As recommended by \citet{momcheva2016}, we limit ourselves to the 23,564 galaxies with $JH_\mathrm{IR}\leq 24$ that are considered to be robust detections and were visually inspected.
    After selecting objects that had at least two separate visual inspections and at $2.0\leq z \leq 2.6$ within the COSMOS field, we are left with 322 objects for cross-correlation from 3D-HST.

\section{Per-survey analysis}
\label{sec:combined-analysis}

    \subsection{Cross-correlation Estimator}
    \label{sec:combined-analysis:xcorr-calc}

        We compute the cross-correlation between the \lya{} forest and galaxy positions using the procedure described in Section 3.4 of \citet{font-ribera-2012}, which we briefly summarize here.
        Given a particular galaxy and its surrounding mean-flux-corrected \lya{} forest fluctuation $\delta_{Fi}$, we sort $\delta_{Fi}$ into bins of distance from the galaxy.
        We bin in both the transverse (which we refer to as $\sigma$) and radial (or $\pi$) dimensions.
        For the transverse $\sigma$ direction, we use 10 bins ranging from 0-30 cMpc, with bin sizes increasing from 0.7 cMpc for the closest two bins to 7.5 cMpc for the furthest two bins.
        For the radial $\pi$ direction, we use 23 bins from -30 to 30 cMpc, again with bin sizes that increase with increasing radial distance.
        For a specific bin $A$, we estimate the cross-correlation within the bin by performing a weighted sum over both all pixels within the bin for a single galaxy, and over all galaxies within the sample:
        \begin{equation}
            \hat{\xi}_A = \frac{\sum_{i \in A} w_i \delta_{Fi}}{\sum_{i \in A} w_i}
        \end{equation}
        The weighting factor $w_i$ is the inverse of the total estimated forest fluctuation variance. This is derived by adding in quadrature the intrinsic variance of the \lya{} forest fluctuation to the instrumental variance.

    \subsection{Error Covariance Estimation}
    \label{sec:combined-analysis:covar}

        We estimate a covariance matrix for the cross-correlation, for each survey in our sample.
        We do this estimation through constructing mock realizations of Lyman-alpha forest surveys and galaxy surveys from an N-body simulation.

        The simulation we use is the Bolshoi-Planck N-body simulation \citep{klypin-2016}, which has a side length of $L=250 \mpch$ and a 2048$^3$ resolution tracking particles of mass $\num{1.5e8}$ \msunh{} each.
        We use the density snapshot and halo catalog at redshift $z = 2.466$. While this differs slightly from the $z=2.30$ midpoint of our $2.0\leq z_\alpha \leq 2.6$ redshift range, the only significant difference is in the mean flux of the Ly$\alpha$ forest. We manually correct this when generating mock Lyman-alpha absorption skewers, as described below.

        To generate \lyaf{} mocks, we first apply the fluctuating Gunn-Peterson approximation (FGPA, \citet{hui-1997}) to the dark matter density field. 
        In the FGPA, the \lya{} optical depth $\tau$ follows a power-law relation with the density $\rho$:
        \begin{equation}
            \tau = \alpha (\rho / \bar{\rho})^\beta
        \end{equation}
        The parameters $\alpha = 0.226, \beta = 1.5$ were chosen by \citet{horowitz-21-tardis-2} to approximately match both observations and cosmological hydrodynamical simulations.
        We then generate flux skewers with sightline density roughly equal to that of CLAMATO's (i.e.\ a mean transverse separation between skewers of 2.4 $\mpch{}$).
        The effects of spectral dispersion were then incorporated by smoothing with a 1D Gaussian kernel of size 1.39 $\mpch{}$, again to match the CLAMATO data.
       Next, we add pixel noise to the skewers using the noise model described in \cite{horowitz-19-tardis-1,horowitz-21-tardis-2}
        as well as an offset to the overall transmitted flux in each skewer to model systematic fitting errors of the skewer's background source continuum \cite{krolewski-17}.

        We divide our simulated skewers into $7 \times 9$ non-overlapping fields over the two dimensions in the Bolshoi-Planck volume that we adopt as the transverse (i.e. sky) plane. Each sub-field sized to match the CLAMATO footprint of 34 $\mpch{}$ $\times$ 28 $\mpch{}$ (assuming a redshift fixed at the mean $z = 2.3$). 
        For the $2 \mpch{}$ overrun in the latter dimension, we take skewers from the other side of the periodic simulation box.
        Since along the radial direction the full redshift range probed by CLAMATO of $2.05 < z < 2.55$ extends beyond the simulation side length of 250 $\mpch{}$, we duplicate all skewers once along the radial direction (which does not introduce flux discontinuities due to the periodic boundary conditions used in the simulation).
        This repetition of the simulation box will cause an underestimate of the cosmic variance contribution to our covariance, but we believe this is likely to be small in comparison with the other sources of error.
        
        For each simulated skewer, we then randomly select one of the real CLAMATO sightlines and apply its pixel mask on to the simulated transmission. 
        This imprints the correct redshift range covered by the observed sightline, as well as masks to account for known intrinsic metal absorption lines in the LBG spectra, broad absorption line (BAL) features in a handful of the quasars, and any damped \lya{} absorbers (DLAs) that were identified in the data.
        
        As mentioned above, the mock flux skewers are computed from a single simulation snapshot at $z = 2.466$, therefore we must correct both for the difference between the simulation redshift and the observational midpoint redshift $z = 2.3$, and for redshift evolution along the skewers.
        We do this by applying a ``mean-flux'' correction.
        From the CLAMATO flux data, we compute the mean flux for redshift bins of size $\Delta z_{\text{bin}} = 0.05$ across the observational redshift range of $2.05 < z < 2.55$.
        Treating the midpoint of the mock skewers as the actual observational redshift $z = 2.3$, we divide the mock skewers into redshift bins of equal size $\Delta z_{\text{bin}} = 0.05$.
        Within each redshift bin, we multiply the mock skewer's flux so that its (weighted) mean flux matches the observational mean flux for the matching redshift bin.

        For each CLAMATO-sized field, we also construct mock redshift surveys from dark matter halos identified within the Bolshoi-Planck volume. 
        For each halo, we use an ``observed stellar mass'' generated from the semi-analytic UniverseMachine code \citep{behroozi-19}, which includes random and systematic observational errors. 
        Like the skewers, we tile the halos through one repetition of the simulation box in the radial direction in order to match the survey volume along the radial direction. %

        For each mock survey, we sample simulated halos following the observed stellar mass distribution. 
        We also only sample from halos within a footprint (relative to the mock CLAMATO field's center) that matches the real survey's approximate outer footprint, but we do not mock-up the detailed CCD gaps or exact mosaic shapes of each survey.
        We stop sampling when the number of halos in the mock survey matches the number of observed galaxies. 
        Finally, we simulate a redshift error and systematic offset by adding a radial shift (in $\mpch{}$ units, not velocity units) drawn from a Gaussian $\mathcal{N}(\delta_{\text{survey},z}, \sigma_{\text{survey},z})$.
        Since we do not \textit{a priori} know the redshift error $\sigma_{\text{survey},z}$ and systematic offset $\delta_{\text{survey},z}$ for each survey, we adopt an iterative procedure for these values, where the fitted redshift error and offset using the mock covariances are fed back to construct new mock surveys, and the fitting procedure repeated until the fitted redshift parameters converge ($<5\%$ difference between iterations).

        Since each CLAMATO-sized footprint in the simulation contains many times more \lya{} skewers and halos than in our observational data, we can construct 3 mock \lya{} surveys out of each footprint by subsampling skewers to match the average transverse intra-skewer separation of CLAMATO. 
        Similarly, we construct 10 mock galaxy redshift surveys within each footprint. For each mock \lya{}-galaxy survey combination, we compute the cross-correlation using the procedure described in Section \ref{sec:combined-analysis:xcorr-calc}. In total, we obtain 63 fields $\times$ 3 \lya{} surveys $\times$ 10 redshift surveys = 1890 mock cross-correlations, computd for each foreground galaxy survey described in Section~\ref{sec:data:gal} .
        The mock cross-correlations use the same 10 transverse and 23 radial bins as the observed cross-correlations, allowing us to compute the $230 \times 230$ covariance matrix $\mathcal{C}$.

    \subsection{Cross-correlation model fitting}
    \label{sec:combined-analysis:xcorr-model}

        We fit each survey's cross-correlation with the linear theory formalism described in \citet{bourboux-20-eboss}, where the Fourier transform of the cross-correlation $\hat{P}_{g,\text{\lya{}}}(\vec{k})$ is given by:
        \begin{equation}\label{eq:pk}
            \hat{P}_{g,\text{\lya{}}}(\vec{k}) = \bgal{} \blya{} (1 + \beta_g \mu_k^2) (1 + \beta_{\text{\lya{}}} \mu_k^2) P_{QL}(\vec{k}) F_{NL} (\vec{k}) G(\vec{k})
        \end{equation}
        The relation between the underlying quasi-linear power spectrum $P_{QL}$ and the galaxy/\lya{} tracers is described by the bias and redshift-space distortion (RSD) parameters $(\bgal{}, \beta_g)$/$(b_{\text{\lya{}}}, \beta_{\text{\lya{}}})$ respectively.
        The systematic galaxy redshift error parameter $\sigz{}$ goes into $F_{NL}$, which is a small-scale correction function, while $G(\vec{k})$ corrects for the effect of cross-correlation binning.
        Finally, before the modeled cross-correlation is compared to the observed cross-correlation, it is offset (before binning) by the systematic redshift parameter $\deltaz{}$.
        In addition, before the cross-correlation in configuration space is compared to the real data, we apply a velocity dispersion resulting from redshift uncertainties from both finite spectral resolution as well as intrinsic uncertainties  in the spectral features used for redshift estimation..
        This is accomplished by convolving the configuration-space cross-correlation with a 1D Gaussian kernel (in the radial direction) of size $\sigz{}$, a free parameter.

        We follow \citet{font-ribera-2012} in assuming a fixed relationship between $\bgal{}$ and $\beta_g$ of $\beta_g = f(\Omega) / \bgal{}$, where $f(\Omega)$ is the logarithmic derivative of the growth rate.
        At $z \sim 2.3$, this is calculable from $\Lambda$CDM as $f(\Omega) = 0.970$, leaving \bgal{} as a free parameter.

        As our observational data set is small, we would be unable to get good constraints on the \lya{} parameters if left free.
        We therefore opt to fix the \lya{} parameters $b_{\text{\lya{}}}$ and $\beta_{\text{\lya{}}}$ to values found by eBOSS \citep{bourboux-20-eboss}.
        However, the values quoted for $b_{\text{\lya{}}}$ and $\beta_{\text{\lya{}}}$ also include the effects of unmasked high column density (HCD) absorbers from the continuum, which can bias the \lya{} parameters.
        Due to the size of our dataset, we find that attempting to similarly include the effect of unmasked HCDs leads to issues in fitting.
        In our fitting process, we therefore turn off all unmasked HCD modeling, and use values of $b_{\text{\lya{}}}$ and $\beta_{\text{\lya{}}}$ found from an analysis of eBOSS data where HCD modeling is also turned off.
        With explicit HCD modeling turned off, the recovered $b_{\text{\lya{}}}$ and $\beta_{\text{\lya{}}}$ values from this eBOSS analysis act as effective bias and RSD parameters, that contain contributions from both the \lya{} forest and HCDs \citep{fontribera-2012-lya-eff-bias}.
        These effective values are $b_{\text{\lya{},eff}} = -0.15$ and $\beta_{\text{\lya{},eff}} = 1.163$.
        To summarize, our free parameters are the galaxy bias $\bgal{}$, the observed galaxy redshift dispersion $\sigz{}$, and the systematic redshift offset $\deltaz{}$.

        We use the \vega{} code \citep{vega, ceceu_2023_vegaexplainer1, dumas_2020_vegaexplainer2} to perform MCMC to obtain constraints on the free model parameters based on our observed cross-correlations, using the covariances obtained from  Section \ref{sec:combined-analysis:covar}.
        We enforce uninformative (flat) priors on all free parameters. 
        As the cross-correlation model we use is based on linear theory, it is ill-suited for small transverse distances due to likely feedback effects (e.g., \citealt{nagamine2021}). 
        We therefore mask out the innermost transverse bins for the data, from 0 to 0.7 $\mpc{}$.
        
        From the sampled posterior, we compute the reduced chi-squared \reducedchisq{} of the maximum a posteriori model, as a measure of whether the best-fit model is overfitting ($\reducedchisq{} < 1$) or underfitting ($\reducedchisq{} > 1$) the data.

    \subsection{Results}
    \label{sec:combined-analysis:results}

        \begin{figure*}
            \centering
            \includegraphics[width=\linewidth]{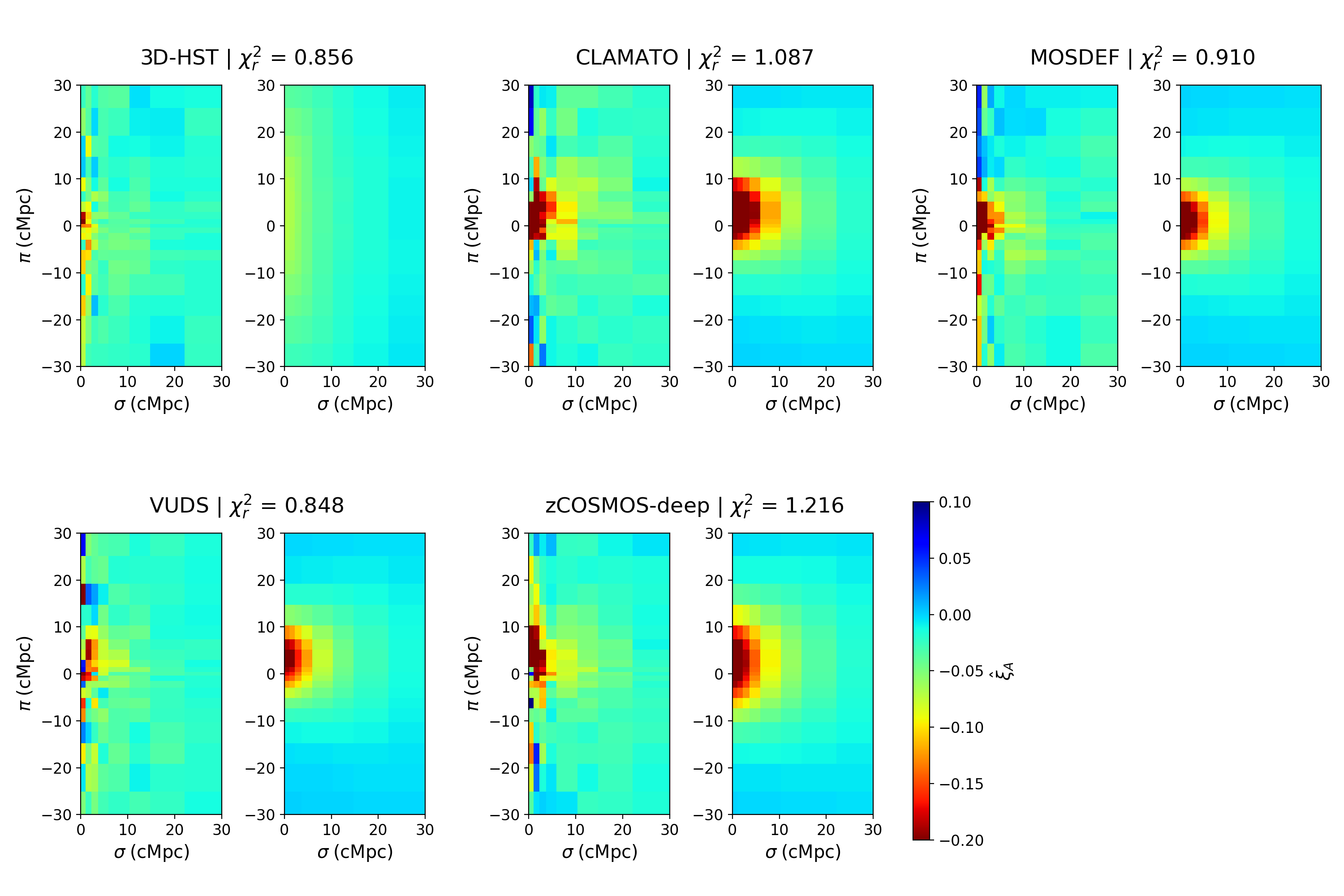}
            \caption{For each redshift survey in the COSMOS field, the left panel of the pair shows the observed cross-correlation between galaxy positions and the \lya{} forest absorption from the CLAMATO survey.
            The right panel of the pair shows the best-fit cross-correlation model using the \vega{} code, and the best-fit reduced chi-squared.
            $\sigma$ is the transverse distance from each galaxy, and $\pi$ is the radial (or line-of-sight) distance.
            Negative $\pi$ values are closer to the observer.
            The transverse bin from 0 to 0.7 cMpc is removed before fitting due to potential feedback effects causing nonlinearity in the cross-correlation, but is still shown in this figure.}
            \label{fig:xcorr-bestfit}
        \end{figure*}

        \begin{figure*}
            \centering
            \includegraphics[width=\linewidth]{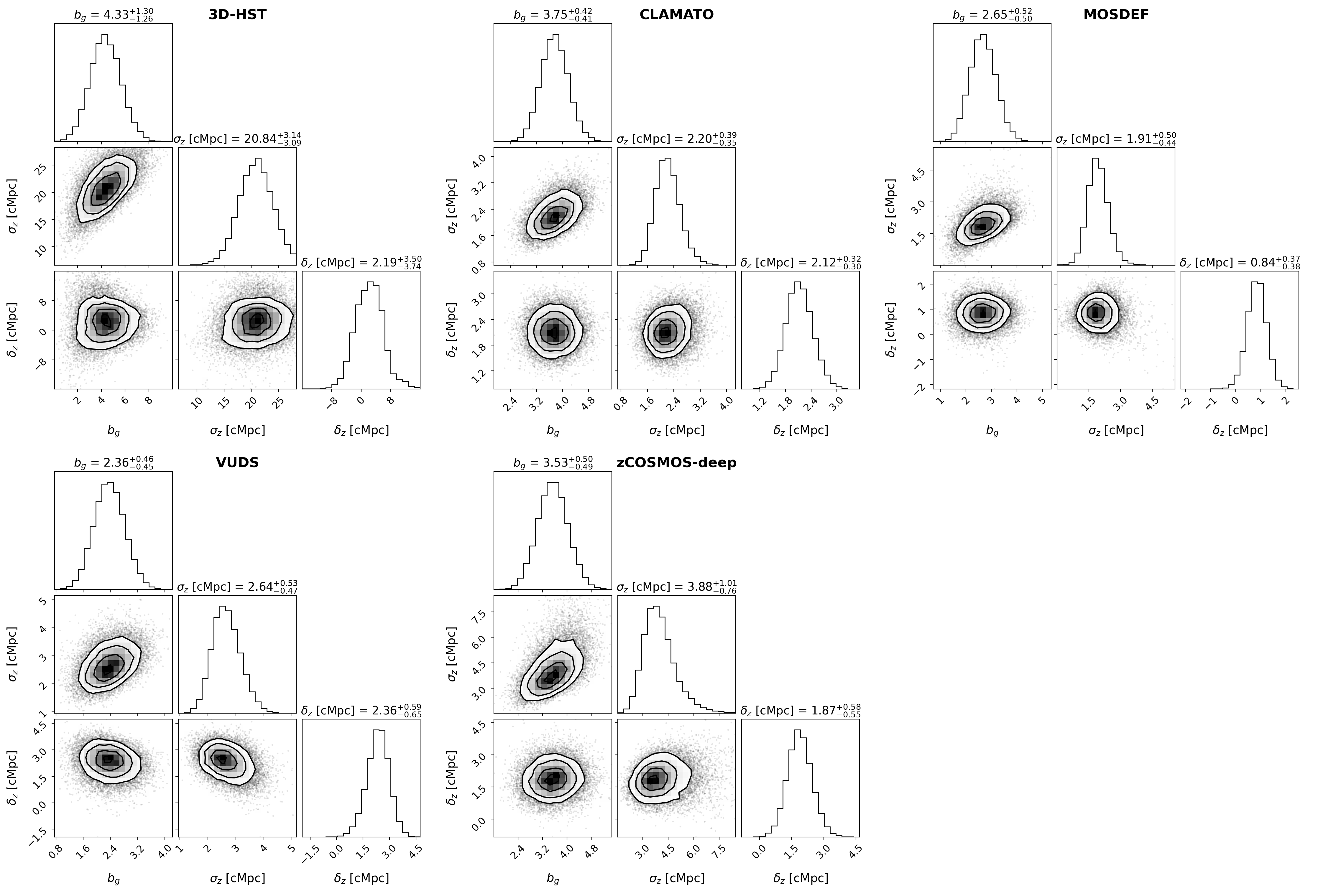}
            \caption{Posteriors on the \vega{} cross-correlation model parameters for each redshift survey, sampled using MCMC.
            The galaxy bias $b_g$ shows a consistent positive correlation with the observed redshift dispersion \sigz{}, so the latter must be carefully treated when connecting the former to halo mass for constraints on the SMHR.}
            \label{fig:xcorr-posteriors}
        \end{figure*}

        For each spectroscopic galaxy redshift survey, we show the observed cross-correlation of their galaxies with the CLAMATO \lyaf{} (left panel) alongside the best-fit \vega{} cross-correlation model (right panel) in Figure \ref{fig:xcorr-bestfit}. 
        A clear negative cross-correlation is seen with all the surveys, which is as expected given that our sign convention for $\delta_F$ is negative for excess absorption. In the case of 3D-HST, however, the signal is noticeably weaker and smeared-out compared to the other samples.
        The radial (i.e. line-of-sight) centroid of some of the cross-correlations also appear displaced in the positive $\pi$ direction, suggesting that the redshifts in the catalogs are underestimated, i.e. they are biased blue-wards. 
        This is particularly noticeable for the the LBG samples (CLAMATO, VUDS, and zCOSMOS-deep).
        
        In comparison with the best-fit models (right panels of Fig.~\ref{fig:xcorr-bestfit}), visually the models appear somewhat discrepant with the observed cross-correlations, but this is due to the significant stochasticity caused by the small number of galaxies ($\sim$200-700) for any one survey.
        To provide a more quantitative comparison, we consider the reduced chi-squared for each best-fit model --- this is labeled at the top of each panel in Figure~\ref{fig:xcorr-bestfit}.
        We find $0.848 \leq \reducedchisq \leq 1.216$ from our best-fit cross-correlation models, 
        for $\nu = 207-3-1=203$ degrees of freedom (recalling that we discard the innermost transverse bins and fit for 3 free parameters).
        These generally indicate reasonable fits, with the range of $\reducedchisq$ bracketing cumulative probabilities of $P = [0.057,0.981]$, respectively, for finding greater \reducedchisq{} values given the degrees of freedom.

        Figure \ref{fig:xcorr-posteriors} shows the MCMC posteriors for the cross-correlation with each galaxy survey.
        In each case, we see a positive degeneracy between \bgal{} and \sigz{}. 
        This is as expected %
        since large galaxy bias values lead to deeper and larger-scale cross-correlation signal, which can also be fitted with a greater line-of-sight dispersion to some extent. 
        This can be seen by comparing the best-fit models for \clamato{} and \mosdef{} in Figure \ref{fig:xcorr-bestfit}, which have similar best-fit \sigz{} but different biases.
        \vuds{} and \zdeep{} also exhibit slight degeneracies between \sigz{} and \deltaz{}, which is not found in the other surveys.
        This may be a result of the comparatively irregular shape of the observed cross-correlation for these two surveys.

        \begin{savenotes}
        \begin{table*}
            \centering
        \renewcommand{\arraystretch}{1.2}
\begin{tabular}{lllll}
\textbf{Survey} & \textbf{$N_s$} & \textbf{$\hat{\delta_z}$ [km/s]} & \textbf{$\hat{\sigma_z}$ [km/s]} & \textbf{Reference $\sigma_z$ [km/s]} \\ \hline
3D-HST & 322 & $+154${\raisebox{0.5ex}{\tiny$^{+247}_{-264}$}} & $1469${\raisebox{0.5ex}{\tiny$^{+222}_{-218}$}} & 1019 \\
CLAMATO & 205 & $+149${\raisebox{0.5ex}{\tiny$^{+23}_{-21}$}} & $155${\raisebox{0.5ex}{\tiny$^{+28}_{-25}$}} & --- \\
MOSDEF & 195 & $+59${\raisebox{0.5ex}{\tiny$^{+26}_{-27}$}} & $135${\raisebox{0.5ex}{\tiny$^{+35}_{-31}$}} & --- \\
VUDS & 469 & $+167${\raisebox{0.5ex}{\tiny$^{+42}_{-46}$}} & $186${\raisebox{0.5ex}{\tiny$^{+37}_{-33}$}} & 150-200 \\
zCOSMOS-deep & 759 & $+132${\raisebox{0.5ex}{\tiny$^{+41}_{-39}$}} & $274${\raisebox{0.5ex}{\tiny$^{+71}_{-54}$}} & ---
\end{tabular}
            \caption{Median values for the redshift dispersion \sigz{} and systematic offset \deltaz{} in velocity units, as well as 16/84th percentile errors. 
            By our convention, a positive value for $\deltaz{}$ implies that on average, a smaller redshift is estimated than the redshift implied by cross-correlation fitting. 
            For comparison, previous estimates for \sigz{} are provided for \threedhst{} \citep{brammer_2012_3dhst}
            and \vuds{} \citep{lefevre_2015_vuds}.}
            \label{tab:redshift-params-velocity}
        \end{table*}
        \end{savenotes}

        Our constraints on \sigz{} and \deltaz{} provide an independent estimate of the observational redshift dispersion and offset.
        We list these estimates converted into velocity units in Table \ref{tab:redshift-params-velocity}, and compare them with previous estimates for \sigz{} where available.
        For the LBG samples (zCOSMOS-deep, VUDS, and CLAMATO), we find redshift errors of order $\sigz \sim 150-250\,\mathrm{km/s}$.
        These are consistent with the literature estimates of the redshift uncertainty, that typically measured using differences across repeated redshift measurements of the same objects, either from within the same survey or with different surveys, as typically reported in the corresponding survey papers (e.g. \citealt{lefevre_2015_vuds} in the case of VUDS).
        The qualitative trends in the comparative LBG redshift uncertainties are also as expected for zCOSMOS-deep, VUDS, and CLAMATO which are all LBG samples: CLAMATO was conducted with a spectral resolution of $R\sim 1100$ and hence has smaller redshift dispersion ($\sigz \sim 160\,\mathrm{km/s}$) than zCOSMOS-deep and VUDS that were both carried out at $R\sim 200$. Of the latter two surveys, VUDS was observed with significantly higher signal-to-noise and thus in turn provides tighter redshifts than zCOSMOS-deep ($\sigz\sim 190\,\kms$ vs $\sigz\sim 250\,\kms$, respectively).

        As for the redshift offsets, all 3 LBG surveys exhibit similar values of $\hat{\delta_z}\approx +150\,\kms$ --- with the positive value indicating that the galaxy redshift estimate is systematically blueshifted relative to the true systemic redshift.  
        This is consistent with the value of $\Delta v = +145^{+70}_{-30}\,\kms$ reported by \citet{rakic:2011} from a comparison of redshifts derived from LBG ISM absorption lines, with high-resolution and signal-to-noise quasar spectra in the KBSS Survey \citep{rudie:2012}. 
        This assumes that the redshifts for CLAMATO, zCOSMOS-deep, and VUDS were derived through the UV absorption lines, and not, e.g. Ly-$\alpha$ emission lines where present. 
        In practice, all three LBG surveys likely have significant fractions of \lya{}-determined redshifts within their samples, with a smaller fraction expected from  
        VUDS which was designed explicitly to achieve generous signal-to-noise on the UV continuum.  
        Given the relatively large uncertainties of the current measurement, we do not pursue this systematic in more detail, but it should be addressed in future analyses with larger samples.

        The redshift error we find for 3D-HST is approximately $\sigz\sim 1400\,\kms$, which is unsurprisingly large given that it is a low-resolution $R\sim 100$ slitless spectroscopic survey. 
        This value is generally consistent ($<2\,\sigma$) with the redshift error quoted by the 3D-HST team \citep{brammer_2012_3dhst}.
        We also find a systematic redshift offset of $\hat{\delta_z}\approx +150\,\kms$  for 3D-HST, albeit in this case consistent with zero given the large uncertainties ($\sigma(\hat{\delta_z}) \sim 250\,\kms$) driven by the lower resolution of the spectra.
        The galaxy bias we derive for 3D-HST is the highest among our different surveys: $b_g = 4.33^{+1.30}_{-1.26}$.
        This supports the fact that 3D-HST is a relatively shallow survey with successful redshifts obtained primarily on brighter or more massive galaxies. 
        However, the errors on the bias measurement from this survey are also $\sim 3\times$ larger than from the other surveys, so we drop 3D-HST from the mass-dependent analysis in the next section.

        Finally, the MOSDEF sample exhibits a redshift scatter of $\sigz \approx 135\,\kms$ which is only marginally smaller than those for the LBG samples. This is surprising, as MOSDEF was carried out with a significantly higher spectral resolution of $R\sim 3000$ than the LBG data. The redshifts were also largely measured from rest-frame optical nebular emission lines detected at high signal-to-noise, therefore one does not expect this to be a significant source of scatter.
        \citet{kriek_2015_mosdef} had reported a $\sigma_\mathrm{nmad}=0.0012$ for the MOSDEF redshift estimates compared with prior redshift estimates, i.e. approximately $100\,\kms$ at $z\sim 2.5$ which is only slightly smaller than what we derive. \citet{rakic:2011}, on the other hand, state that the expected statistical uncertainty on redshift estimates is of order $\sim 60\,\kms$. 
        While this deviation in the fitted MOSDEF $\sigz{}$ is only at the 2.4$\sigma$ level, one
        possibility for the increased $\sigz$ is the effect of fingers-of-god (FoG) within virialized halos. 
        In our analysis, we have chosen to interpret the small-scale $F_{NL}$ term in Equation~\ref{eq:pk} as arising largely from redshift errors. This is because we expect the FoG to generally be a small effect because only $<10\%$ of galaxies at these survey depths are expected to be satellites (e.g. \citealt{ishikawa:2017}).  
        However, the MOSDEF footprint within COSMOS is now known to intersect multiple overdensities \citep{ata:2022} within the relevant redshift range despite its small footprint, notably at $z\approx 2.11$ \citep{nanayakkara2016}, $z\approx 2.30$ \citep{lee-2016-clamato-dr1-protocluster,dong_2023_costco}, and $z\approx 2.45-2.55$ \citep{casey2015,wang2016,cucciati2018}.
        Our MOSDEF sample therefore likely has a significant contribution from satellite galaxies orbiting galaxy group-sized halos within the nascent protocluster cores, leading to non-negligible FoG contributions to our inferred $\sigz$.
        On the other hand, the systematic redshift offset we determine for MOSDEF is $\hat{\delta_z}\approx +59^{+26}_{-27}\,\kms$. This is marginally consistent ($\sim 2\,\sigma$) with zero, which is the expected value since nebular emission lines are expected to accurately trace the galaxy systemic redshift.

        As for the galaxy bias parameters $\bgal{}$, we find values that are inconsistent at a $1-2\sigma$ level across the different galaxy surveys.
        As a larger bias value implies that the galaxies reside in environments with more \lya{} absorption and therefore higher matter density, these different bias values indicate that 
        the galaxy samples for the surveys likely represent different halo masses.
        In the next section, we use this fact to obtain constraints on the stellar-halo mass relation.

\section{Mass-dependent analysis}
\label{sec:split-analysis}

    \begin{figure*}
        \centering
        \includegraphics[width=0.9\linewidth]{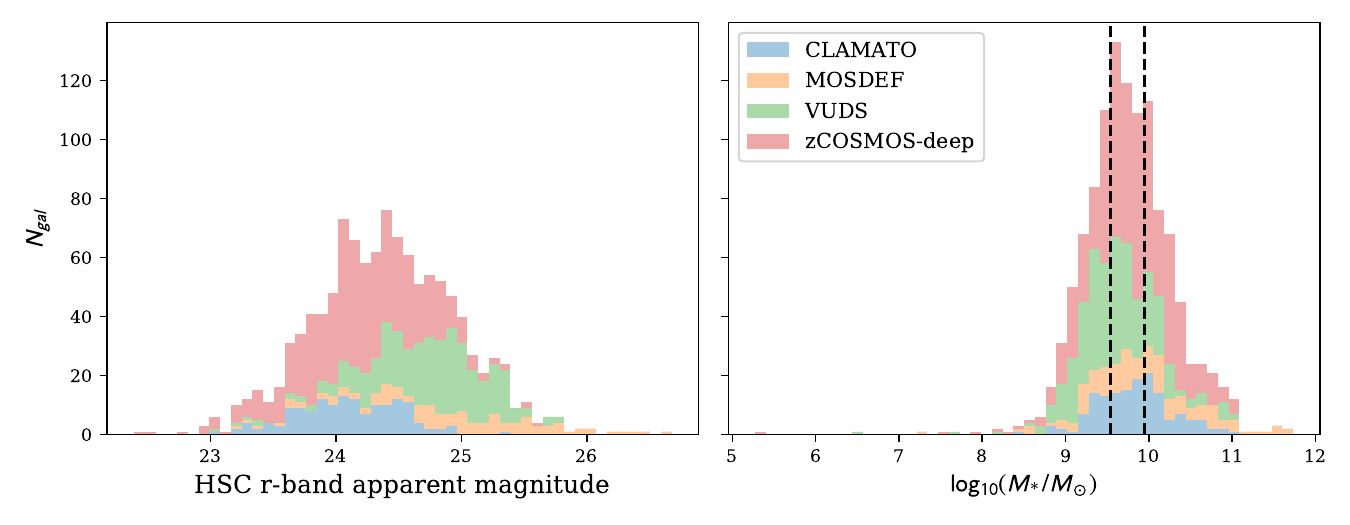}
        \caption{Stacked histogram of Hyper Suprime-Cam r-band apparent magnitude and stellar mass, for surveys used in our mass-dependent analysis. Both quantities are from the COSMOS2020 Farmer catalog \citep{weaver_2022_cosmos2020}. 27 galaxies in VUDS and MOSDEF have no matching COSMOS2020 photometry, and are excluded from the left panel. For our mass-dependent analysis, we split each survey (which contains only unique galaxies) into 3 stellar-mass bins.
        These splits, which are the 33rd and 66th percentiles of the combined stellar mass distribution, are the black dotted lines, and occur at $\log_{10}(M_* / \msun{}) = 9.538$ and $\log_{10}(M_* / \msun{}) = 9.948$.}
        \label{fig:split-gal-sample}
    \end{figure*}

    As discussed in the Introduction, \citet{newman-24-latis} use galaxy-Ly$\alpha$ forest cross-correlation from the LATIS survey to infer the average halo mass of galaxy samples across several mass bins, and provide constraints on the stellar mass-halo mass relationship at $z \sim 2.5$.
    We carry out a similar analysis with the galaxy surveys in the COSMOS field coeval with the CLAMATO \lyaf{} data, but with some significant differences in our approach.
    While \citet{newman-24-latis} directly fit the cross-correlation for each binned sample to an average halo mass parameter using synthetic cross-correlations computed from the ASTRID hydrodynamic simulation, we carry out the intermediate step of first fitting for an galaxy bias value to different stellar mass bins, and then connect the bias to an average halo mass using the Bolshoi-Planck N-body simulation.

    While our aggregate galaxy sample in the COSMOS field is of comparable size to LATIS (albeit slightly smaller), we must address the fact that our sample is comprised of several different galaxy survey samples that each have a different observed redshift dispersion $\sigz{}$ and systematic redshift offset $\deltaz{}$. These must be separately accounted for in the analysis.

    To produce our galaxy sample for this analysis, we make a unique selection of galaxies to avoid having galaxies that are represented in the different redshift surveys. This uses the following selection priority for the redshifts, in descending order: MOSDEF, CLAMATO, VUDS, zCOSMOS-deep.
    As previously mentioned, we discard all galaxies with redshifts only measured by \threedhst{} due to its significantly larger error bars on the galaxy bias.
    Thanks to the rich multi-wavelength photometric data available in the COSMOS field, it is possible to obtain stellar mass estimates that are based on robust NIR photometry by cross-matching with the COSMOS2020 Farmer photometric catalog \citep{weaver_2022_cosmos2020}.
    This cross-matching largely follows \citet{momose:2024}, and we drop galaxies which have no estimated stellar mass from the COSMOS2020 catalog.

    As COSMOS2020 estimates stellar mass jointly with the photometric redshift, catastrophic failures to fit the photometric redshift for faint galaxies can bias our stellar mass, especially on the low end.
    To guard against this, we remove 39 galaxies which have a estimated photometric redshift $z_{\text{phot}} < 1$.
    These are mostly from VUDS, which has 26 galaxies meeting this criterion.

    We note that our sample includes 6 galaxies with $\log_{10}(M_* / \msun{}) < 8$, which is below the stellar mass completeness threshold quoted in \citet{weaver_2022_cosmos2020}.
    Since this is a comparatively small fraction of our sample, we do not remove it from the subsequent analysis.
    Also, because this is a \textit{cross}-correlation measurement, we do not need to correct for incompleteness in the lower-mass bins as is the case for auto-correlation measurements.
    
    The final size of our combined dataset is 1150 galaxies, from the original 1950 (non-unique) galaxies used in our per-survey analysis.
    We then split our aggregate galaxy sample (shown in Figure \ref{fig:split-gal-sample}) into three equally-sized stellar mass bins, at the 33rd and 66th percentiles which we refer to as the `low', `mid' and `high' stellar mass bins.
    These bins are separated at $\log_{10}(M_* / \msun{}) = 9.538$ and $\log_{10}(M_* / \msun{}) = 9.948$, such that the median log-stellar masses are $\mathrm{med}(\log_{10}(M_* / \msun{})) = [9.28, 9.74, 10.22]$ respectively.
    
    We note that the VUDS survey is over-represented in the lowest stellar-mass bin. Since these low-mass galaxies might be expected to be the faintest and thus have lower signal-to-noise, we need to guard against the possibility that the quality of the spectroscopic redshifts might be systematically lower in the low-mass bin. 
    Checking the visual inspection flags in the VUDS DR2 (Tasca et al, in prep) sample, we find that  galaxies flagged as `2' (75-85\% probability to be correct per \citealt{lefevre_2015_vuds}) --- the lowest quality we use --- are roughly equally distributed across the mass bins. 
    We are thus confident that the quality of the spectroscopic redshifts are robust in VUDS even in the low-mass sample.

    For all galaxies in a given survey that fall in the stellar mass bin, we compute the observed cross-correlation using the same procedure as Section \ref{sec:combined-analysis:xcorr-calc}.
    We approximate the covariance for each stellar mass bin by rescaling the covariance calculated in Section \ref{sec:combined-analysis:covar} by $N_{\text{bin}} / N_{s}$, where $N_{\text{bin}}$ is the number of survey galaxies within the mass bin, and $N_s$ the total number of survey galaxies.

    We then fit these samples' cross-correlations using the \vega{} code to the cross-correlation model described in Section \ref{sec:combined-analysis:xcorr-model}, except that we fit a separate galaxy bias for each stellar mass bin.
    We leave the observed redshift dispersion $\sigz{}$ and systematic redshift offset $\deltaz{}$ as free parameters that are common to all mass bins in a survey, as these arise from instrumental effects and the common population within the survey, and to zeroth order we do not expect these parameters to depend on the bias or stellar masses of the galaxies.
    In summary, the free parameters for which we sample the posterior using MCMC are: the galaxy biases in the `low', `mid', and `high' stellar mass bins $b_{g,lm}, b_{g,mm}, b_{g,hm}$, the redshift dispersion $\sigz{}$, and the systematic redshift offset $\deltaz{}$.

    We additionally follow \citet{newman-24-latis} in performing a ``2-dimensional'' fit, where we sum the cross-correlation along all bins in the transverse direction for both the model and data.
    This largely obviates the effects of the redshift error and offset, as these apply only in the radial (line-of-sight) direction\footnote{Although we note that the systematic offset $\deltaz{}$ still has a second-order effect on the \projd{} cross-correlation due to the finite total extent of our radial cross-correlation bins. Because of this, in our \projd{} analysis we fix \deltaz{} and \sigz{} to the best-fit values from our \alld{} mass-dependent analysis.}.
    This provides an additional consistency check on our mass-split fits.

        \begin{figure*}
            \centering
            \includegraphics[width=0.7\linewidth]{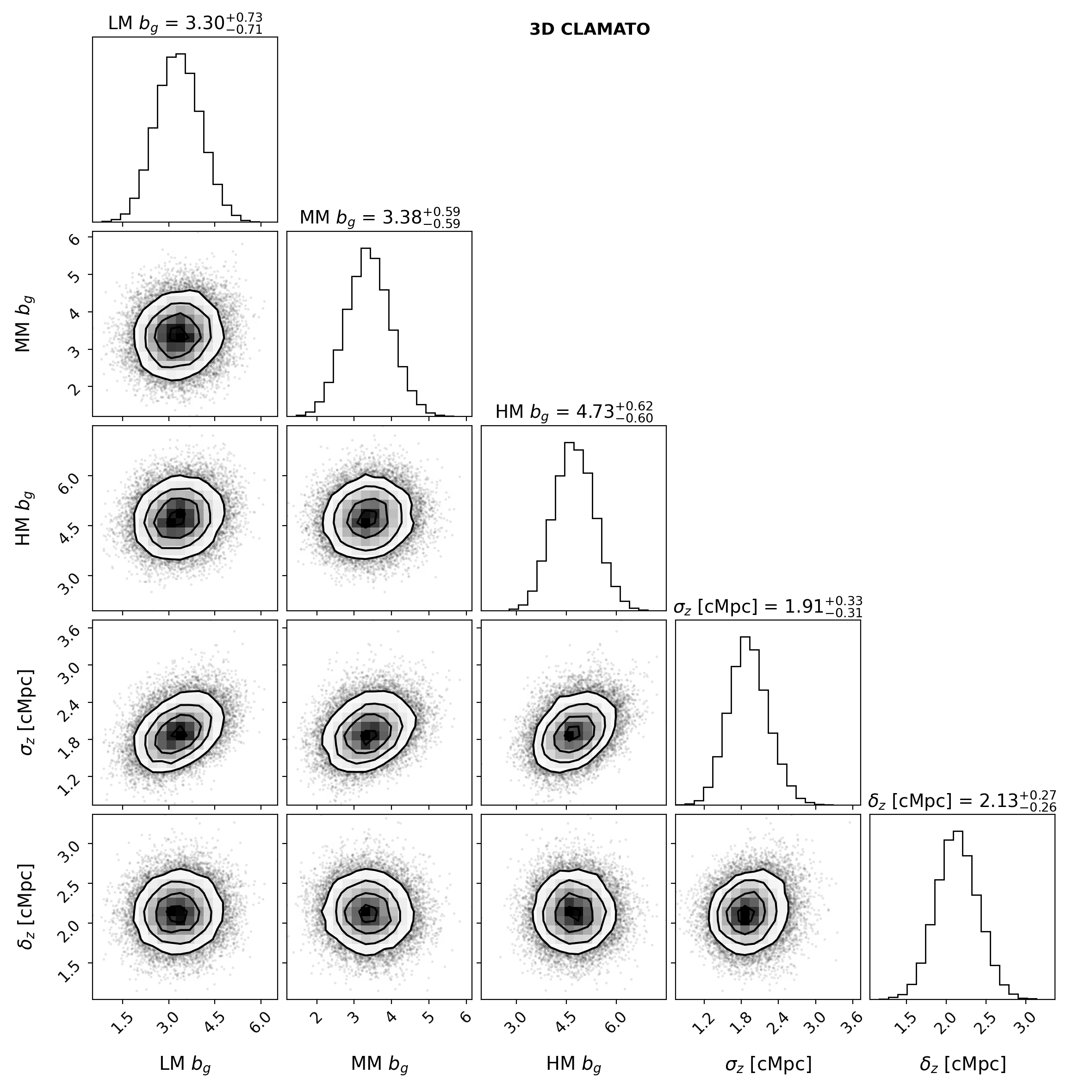}\\[2em]
            \includegraphics[width=0.45\linewidth]{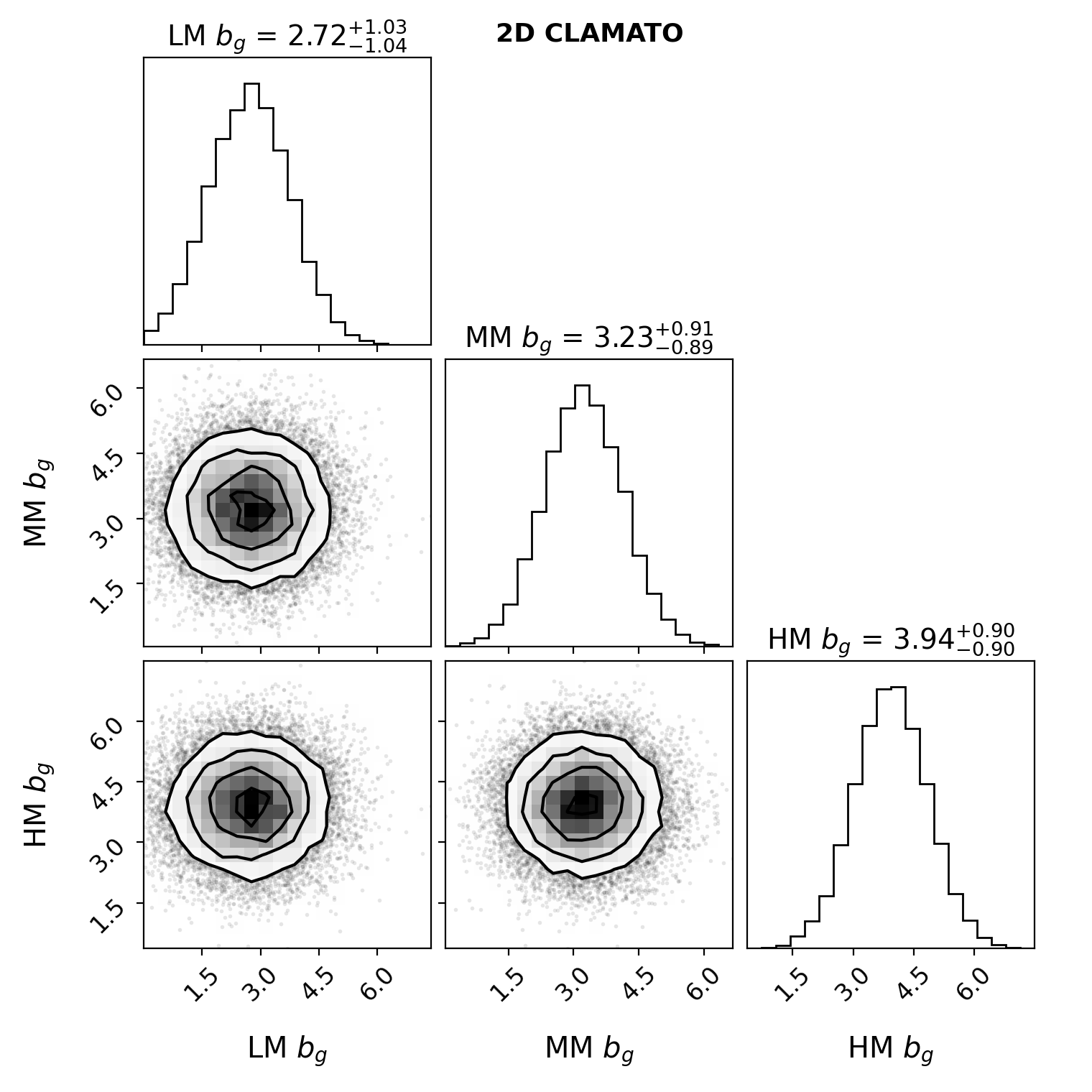}
            \caption{Selected example of a mass-dependent model posterior, for the CLAMATO galaxy redshift survey. The ``\alld{}'' analysis where the model is fit to the full transverse + radial cross-correlation is shown at the top, as well as the ``\projd{}'' analysis where the model is fit to a radially-summed cross-correlation (bottom).
            One galaxy bias is fit per stellar mass bin; for the \alld{} analysis, the redshift dispersion/offset parameters \sigz{} \& \deltaz{} are shared across all stellar mass bins.
            Since summing the cross-correlation in the radial direction almost completely eliminates the effect of the redshift parameters, they are not fit in the \projd{} analysis.
            \label{fig:example-split-corner} }
        \end{figure*}

        We show the posteriors obtained from the \alld{} and \projd{} mass-dependent analysis for CLAMATO in Figure \ref{fig:example-split-corner}.
        For all the galaxy surveys, we find that the redshift dispersion and systematic offset obtained from the mass-dependent analysis is consistent with the results from the per-survey analysis on a 1$\sigma$ level.
        This is as expected, as the redshift parameters are shared across all three mass bins, such that the redshift parameters are constrained by exactly the same galaxy sample as the per-survey analysis.
        The $<1\sigma$ differences in the redshift parameter posteriors between the per-survey and mass-dependent analyses are likely due to the procedure we use to obtain the covariance for each mass bin.
        This procedure involves scaling the per-survey covariance by the number of galaxies in each bin, therefore the rescaled covariance is unlikely to match the ``true covariance'' for each mass bin.
        This is because $\bgal{}$ and $\sigz{}$ actually have a positive correlation, and the $\sigz{}$ used for mock galaxy survey generation for the covariance is the value fit from the per-survey analysis, with its corresponding bias.

        \begin{figure*}
            \centering
            \includegraphics[width=\linewidth]{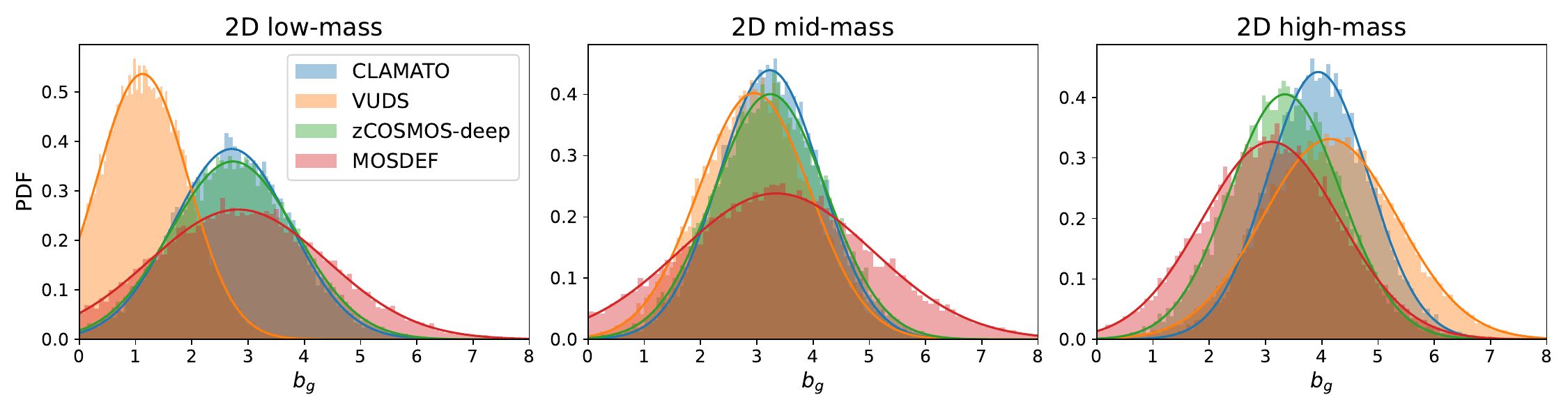}\\
            \includegraphics[width=\linewidth]{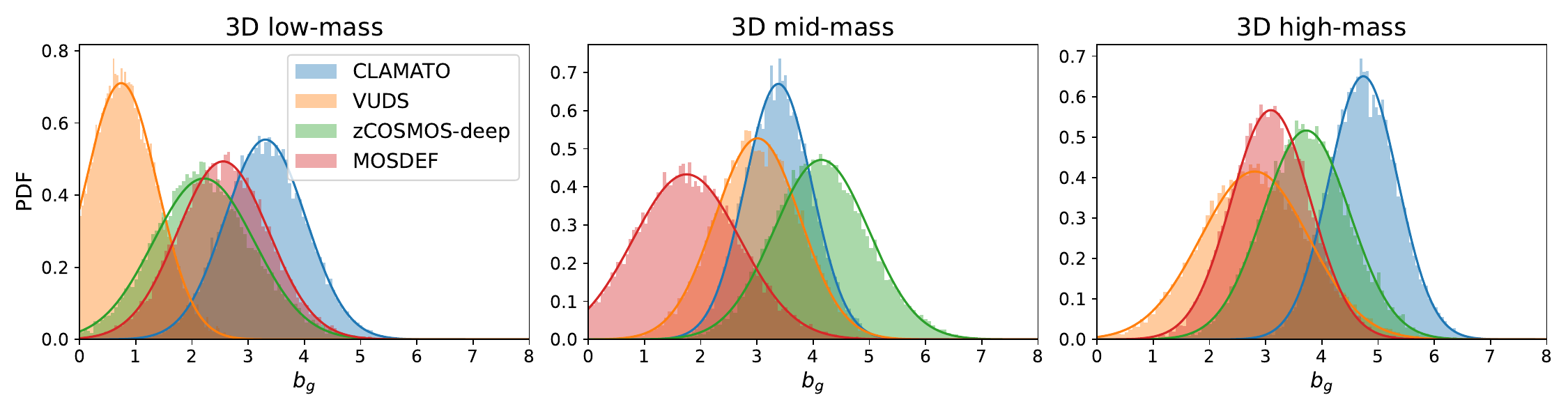}
            \caption{Marginalized mass-dependent posteriors on the galaxy bias for all redshift surveys we use in our mass-dependent analysis.
            The MCMC-sampled posteriors are shown as histograms, and a fitted zero-truncated Gaussian is overlaid on top.
            While the expected trend of higher stellar-mass bins having higher biases (and therefore higher average halos masses) is generally followed, the \alld{} mid-mass bias for MOSDEF is a notable exception.
            The abnormally-low bias fitted for that bin possibly arises from hotter protocluster gas caused by AGN feedback \citet{dong:2024}, with 8 out of 34 galaxies in the bin being $<10$ Mpc from the protoclusters.}
            \label{fig:bias-posteriors-split} 
        \end{figure*}

        Figure \ref{fig:bias-posteriors-split} shows the recovered posteriors for $\bgal{}$ for each stellar mass bin and each survey, with accompanying best-fit Gaussians (truncated at $\bgal{} = 0$).
        The posteriors from the \projd{} (transverse-summed cross-correlation) analysis are noticeably wider than the corresponding \alld{} analysis, which is expected given the loss of information when summing along the transverse direction.

        While the recovered $\bgal{}$ from the individual surveys generally increase for the higher stellar mass bins, a notable exception exists for the \alld{} analysis of \mosdef{}.
        As shown in the bottom panel of Figure \ref{fig:bias-posteriors-split}, the recovered mid-mass bias for this bin is significantly ($\sim1 \sigma$) lower than the low- and high-mass bins.
        All other surveys for the \projd{} and \alld{} analyses have an approximately monotonically increasing bias with higher mass bins as expected, so this \mosdef{} bin deserves a closer look.
        While this may just be the result of cosmic variance with only 34 galaxies within the mass bin, we believe this effect may also come from the presence of heated proto-ICM from one of the protoclusters within the survey volume \citep{dong:2024}. 
        8 of the 34 MOSDEF galaxies within this mass bin are $<10$ \mpc{} from the COSTCO-I system found by \citet{dong_2023_costco} (see also \citealt{lee-2016-clamato-dr1-protocluster}).
        This protocluster was found from observations to exhibit an anomalous lack of \lya{} absorption within a radius of $\sim 13$ \mpc{} ($\sim 4$ pMpc at $z \sim 2.3$), which would correspondingly suppress the fitted bias for galaxies within this radius.
        However, it is unclear whether this same suppression is also manifested in the \projd{} analysis due to the relatively larger uncertainties in the fit.

        \begin{figure*}
            \centering
            \includegraphics[width=\linewidth]{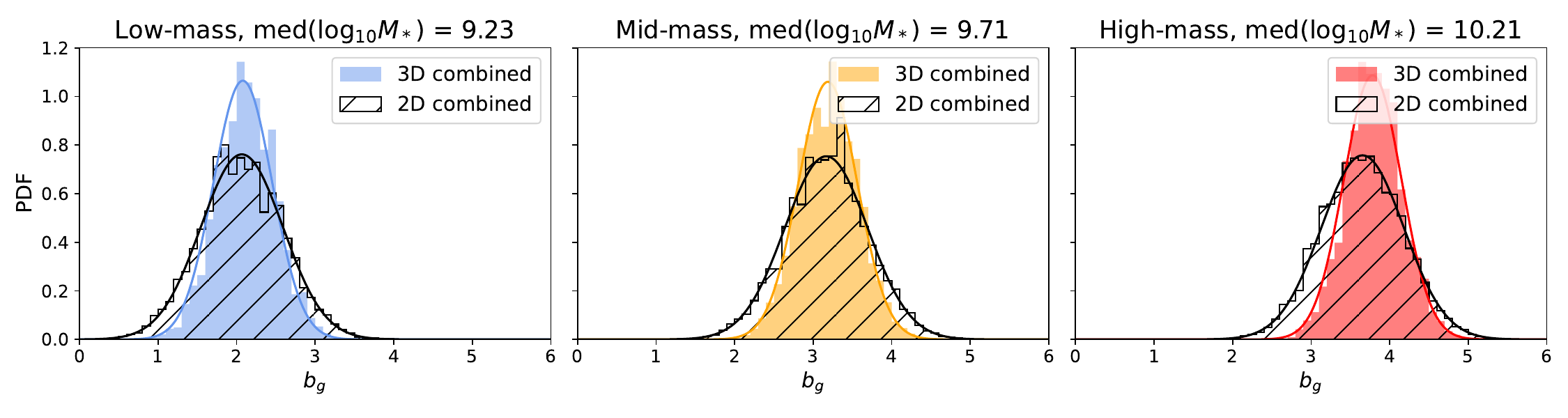}
            \caption{Combined galaxy bias posteriors for each stellar mass bin.
            For both \alld{} and \projd{} analyses, we combine the posteriors by multiplying the probability density distributions of Gaussians fit to the marginalized bias posteriors (see Figure \ref{fig:bias-posteriors-split}).
            This yields the probability density distribution of another Gaussian (the solid line).
            We also multiply the density histogram bins of the marginalized posteriors, which is shown as the histogram bins in the figure.
            While this procedure ignores covariances between biases in different mass bins for each redshift survey, we find that these covariances are generally small ($r < 0.1$).}
            \label{fig:bias-posteriors-combined}
        \end{figure*}

        Once we obtain MCMC posteriors for the galaxy biases within the individual stellar mass bins for each survey, we combine them across the different surveys to get a final posterior on the galaxy bias for each stellar mass bin.
        This assumes that the bias measurements are independent across surveys.
        Additionally, we assume that the covariance between $b_{g,lm}, b_{g,mm}$, and $b_{g,hm}$ measured within the survey can be neglected.
        As the correlation coefficients we find between the biases within the same survey are $<0.1$, this is generally true.
        We fit a truncated Gaussian with a lower limit of 0 to the posterior distribution for the bias estimated for each stellar mass bin in the individual surveys.
        Across the different surveys, we then multiply the posterior PDFs of the fitted Gaussians for each stellar mass bin.
        This produces a final truncated Gaussian PDF, which we take as the combined galaxy bias posterior for that stellar mass bin.

        The result of combining the bias posteriors across all surveys for each mass bin is shown in Figure \ref{fig:bias-posteriors-combined}.
        In addition to multiplying the fitted Gaussian PDFs, we also show combined histogram bins.
        To obtain these combined bins, we construct posterior density histograms using a common set of bias bins for all surveys, then multiply the density in each bias bin and renormalize.

        For our stellar-mass sub-samples with $\lgmstar = [9.28, 9.74, 10.22]$, we find $b_g = [2.07^{+0.52}_{-0.52}, 
        3.17^{+0.53}_{-0.53}, 3.65^{+0.52}_{-0.52}]$ from the 2D analysis and 
        $b_g=[2.08^{+0.37}_{-0.37}, 3.19^{+0.37}_{-0.37}, 3.78^{+0.36}_{-0.36}]$ from the 3D analysis, respectively.
        These results are also tabulated in Table~\ref{tab:bias_mh}, while the full Gaussian fit parameters for $\bgal{}$ for the mass-dependent analysis in all the surveys are available in Appendix \ref{appendix:mass-split-table}.
        
        Despite the large variance in bias fits across surveys in each stellar mass bin, the \projd{} and \alld{} combined bias posteriors are consistent with each other to well within $1\sigma$.
        The \alld{} results do have better precision, as expected given that the projection done in the \projd{} discards some information.
        In particular, there is a differentiation of the galaxy biases between the high- and low-mass bins at the $\sim 4\sigma$ level from the \alld{} results, while in the case of the \projd{} results the differences are at the $\sim 3\sigma$ level even though the monotonic increase of galaxy bias with stellar mass is still apparent.

\begin{table*}
\centering
\renewcommand{\arraystretch}{1.8}
\begin{tabular}{lcccccc}
\hline
\multirow{2}{*}{Mass Bin} & \multirow{2}{*}{$\mathrm{med}\left(\log_{10} \left(\frac{M_*}{M_\odot}\right) \right)$} & \multicolumn{2}{c}{\projd}                                            &           & \multicolumn{2}{c}{\alld}                                              \\ \cline{3-4} \cline{6-7} 
                          &                                                                                         & $b_g$                  & $\log_{10} \left(\frac{M_h}{M_\odot}\right)$ & \phantom{i} & $b_g$                  & $\log_{10} \left(\frac{M_h}{M_\odot}\right) $ \\ \hline
Low Mass                  & 9.28                                                                                    & $2.07^{+0.52}_{-0.52}$ & $10.44^{+0.67}_{-1.39}$                      &           & $2.08^{+0.37}_{-0.37}$ & $10.47^{+0.51}_{-1.37}$                       \\
Mid Mass                  & 9.74                                                                                    & $3.17^{+0.53}_{-0.53}$ & $11.63^{+0.36}_{-0.44}$                      &           & $3.19^{+0.37}_{-0.37}$ & $11.65^{+0.24}_{-0.30}$                       \\
High Mass                 & 10.22                                                                                   & $3.65^{+0.52}_{-0.52}$ & $11.95^{+0.38}_{-0.36}$                      &           & $3.78^{+0.36}_{-0.36}$ & $12.06^{+0.24}_{-0.27}$                      
\end{tabular}
\caption{\label{tab:bias_mh}
    Galaxy bias and host halo masses, inferred from the combined survey sample split into 3 equal bins in stellar mass. All quoted values for the halo mass use the \vega{}-derived bias-halo mass relation.
    All central values in this table are the median, and listed errors are the 16/84th percentiles.}

\end{table*}

    \subsection{Connecting galaxy bias to halo mass}
    \label{sec:split-analysis:bias-hmass}

        Now that we have a measurement of the galaxy bias for each stellar mass bin, we still need to derive a mean halo mass for each bin in order to place constraints on the stellar mass-halo mass relation.

        We first attempt to do this using the formal theoretical definition of the galaxy bias mass-selected subsamples of halos from the Bolshoi-Planck N-body simulation.
        Separating the halos by halo mass into bins of size 0.1 in $\log_{10}(M_h)$, we do the following for each bin.
        We use the \texttt{nbodykit} code to compute the halo power spectrum up to $k_{\text{max}} = 0.2$, and calculate the theoretical mass-dependent galaxy bias from its definition in linear theory,
        \begin{equation}
            b_{g,\text{th}}(\overline{M}_h) = \text{med}_k(\sqrt{P_h(k) / P_L(k)}) \;,
        \end{equation}
        where $\overline{M}_h$ is the mean halo mass of the bin, $P_h(k)$ and $P_L(k)$ are the halo and linear power spectra respectively, and $\text{med}_k(\cdot{})$ denotes taking the median bias over $k$ up to the calculated maximum of 0.2.
        Implicitly, in this procedure we are assuming that each halo is occupied by exactly one galaxy.
        This is necessitated by our usage of the halos to construct mock covariances in Section \ref{sec:combined-analysis:covar}, but is reasonable in light of observational auto-correlation measurements of LBGs at $2<z<5$ that imply satellite fractions of $<10\%$ (e.g. \citealt{ishikawa:2017}).

        However, in our observations we find that significant correlations exist between \bgal{} and \sigz{}, which can be seen in Figure \ref{fig:xcorr-posteriors}. Even if marginalized over all other parameters, the \bgal{} posterior distributions we obtain by fitting \vega{} to our data will differ from the theoretical bias value, depending on the best-fit \sigz{}.
        In order to more realistically propagate the combined bias posterior from all surveys to a halo mass estimate, we use the mock cross-correlation pipeline we construct in Section \ref{sec:combined-analysis:covar} to estimate the cross-correlation covariance.

        Using the best-fit \sigz{} and \deltaz{} we have determined for each survey, we draw 1000 bootstrap realizations of the redshift offset of the combined-survey galaxy sample, using our mock galaxy catalog.
        From the bootstrap realizations, we fit a combined \sigz{} and \deltaz{}, which we then use to create a set of halo mass-binned galaxy mocks.
        For 44 halo mass bins equally spaced in $\log_{10}M_h$, from $\log_{10}M_h = 9$ to $\log_{10}M_h = 13.4$, we create 1000 galaxy mocks for each bin.
        Combined with the mock \lya{} skewers generated for the covariance estimation, we create a set of halo-mass binned cross-correlations, 44000 in all.
        For each of the cross-correlation realizations, we fit \vega{} to it (using likelihood maximization, not MCMC, for speed).
        We use the same model assumptions described in Section \ref{sec:combined-analysis:xcorr-model}, except that we fix the \sigz{} and \deltaz{} \vega{} parameters to the values used to generate the mocks.
        We use the mock-based covariances, doing the same galaxy-number rescaling procedure as the actual mass-dependent analysis.

        \begin{figure*}
            \centering
            \includegraphics[width=0.5\linewidth]{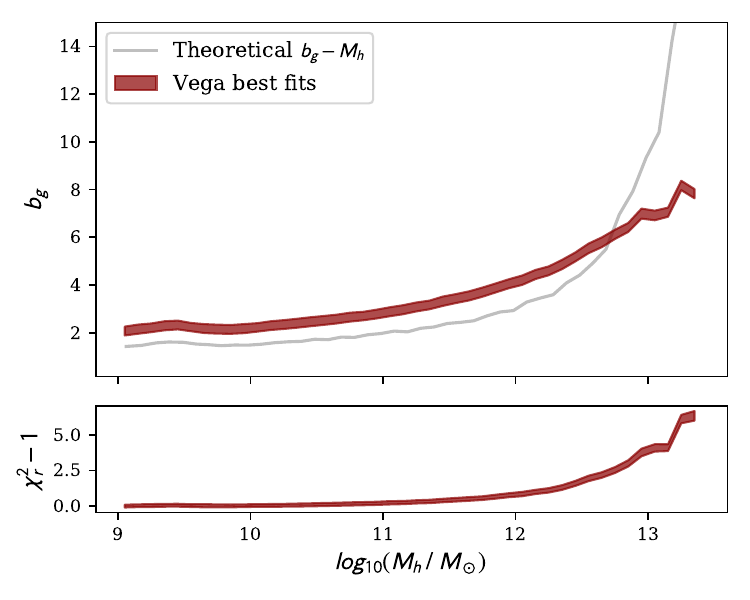}
            \caption{Galaxy bias-halo mass relation, fit using two strategies. The grey line calculates the relation from the Bolshoi-Planck N-body simulation, using the formal definition of linear galaxy bias along with the halo and linear matter power spectra. 
            The dark red line calculates the relation by constructing halo mass-binned mock cross-correlations from Bolshoi-Planck, and fitting the \vega{} code to them with a nonzero redshift dispersion \sigz{}.
            As multiple mock cross-correlation realizations exist per halo mass bin, the thickness of the line denotes the 1$\sigma$ errors on the fitted bias.
            The residual in the reduced chi-squared $\chi_r^2$ of the \vega{} best fit is shown on the bottom inset.}
            \label{fig:bias-halomass-curve}
        \end{figure*}

        We plot the theoretical bias-halo mass curve in Figure \ref{fig:bias-halomass-curve}, as well as the mock-based, \vega{}-fit curve.
        At the bottom of the figure, we plot \vega{}'s best-fit reduced chi-squared.
        As a result of the breakdown in linear theory, the halo bias begins to diverge at the high halo mass end ($M_h > 10^{12} \msun{}$).
        However, since we have no galaxies above this mass in our sample this does not pose an issue.
        Due to the non-zero \sigz{} used when fitting \vega{}, we see that \vega{} estimates a higher bias than the theoretical curve for the halo mass range we care about ($M_h < 10^{12} \msun{}$).
        \vega{} also begins to diverge in the high-mass regime due to the close-by cross-correlation becoming non-linear, which is reflected in the increasing best-fit $\chi_r^2$ shown in the bottom inset.

        When propagating our combined bias posteriors to constraints on the SHMR, we use the \vega{}-derived relation.
        An important assumption of this approach is that we treat the bias-halo mass relation as deterministic, taking the median bias of the \vega{} curve.
        This means we ignore inherent stochasticity in the galaxy bias-halo mass relation, but a full treatment of the stochastic nature of the relation is outside the scope of this work.

    \subsection{Stellar-halo mass relation constraints}
    \label{sec:split-analysis:shmr}

        \begin{figure}
            \centering
            \includegraphics[width=\linewidth]{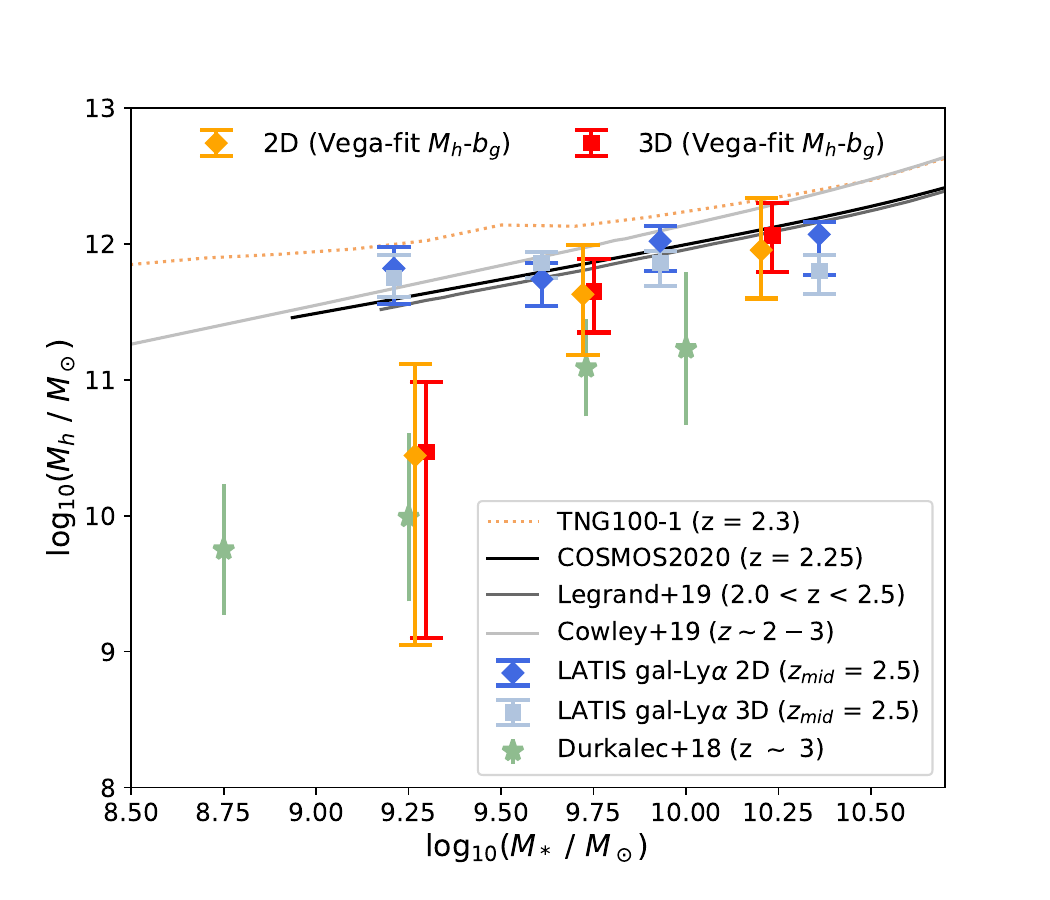}
            \caption{Constraints on the stellar-halo mass relation (SMHR).
            Orange and red error bars are results from our 2D and 3D cross-correlations analyses, respectively. The median log-stellar masses for the 2D and 3D analyses are visually shifted by $\pm0.015$ for clarity.
            We compare our results to similar constraints from galaxy-\lya{} cross-correlations on the LATIS \lya{} survey \citep{newman-24-latis}, as well as $z \sim 3$ constraints from VUDS galaxy-galaxy clustering \citep{durkalec_2018_vuds_clustering}.
            We also include observational constraints derived from angular correlation measurements \citep{cowley_2019_z23_shmr, shuntov_2022_cosmos20_shmr} and abundance matching \citep{legrand_2019_cosmos15_shmr}.
            For \citet{cowley_2019_z23_shmr}, the included curve is only for star-forming galaxies.
            The dotted curve indicates predictions of the SMHR from the IllustrisTNG hydrodynamical simulation \citep{nelson_2019_tng}
            \label{fig:hmass-smass-relation}}
        \end{figure}

        After propagating our bias posteriors through the $\bgal{}$-$M_h$ relation, we obtain constraints on the stellar-halo mass relation, which are shown in Figure \ref{fig:hmass-smass-relation} and tabulated in Table \ref{tab:bias_mh}.
        For comparison, we show similarly derived constraints from the galaxy-\lya{} cross-correlation in the LATIS \lya{} forest survey \citep{newman-24-latis}, as well as spectroscopic galaxy-galaxy clustering constraints at $z \sim 3$ from the VUDS survey \citep{durkalec_2018_vuds_clustering}.
        We additionally include observational constraints derived from angular auto-correlation measurements using the COSMOS2020 catalog \citep{shuntov_2022_cosmos20_shmr} and the SMUVS survey \citep{cowley_2019_z23_shmr} catalogs.
        For the latter, the SHMR is calculated for central galaxies in a single redshift bin spanning $z \sim 2-3$, divided into passive and star-forming populations.
        As our total observational sample is dominated by star-forming galaxies' redshifts (nebular line-emitting galaxies for MOSDEF and LBGs for the other surveys), we compare only with the star-forming results from \citep{cowley_2019_z23_shmr}.
        Finally, we show the SMHR predicted from the Illustris TNG TNG100-1 hydrodynamical simulation, where
        we follow the methodology of \citet{shuntov_2022_cosmos20_shmr} for calculating the stellar and halo masses of each galaxy in the simulation.

        The \alld{} halo mass constraint we obtain for our smallest stellar mass bin (median log stellar mass of 9.28) is more consistent with the results of \citet{durkalec_2018_vuds_clustering} than with angular correlation/abundance matching observations or the LATIS results.
        However, while a $\sim 2.5\sigma$ discrepancy exists between our observations and the latter, we caution that for this smallest stellar mass bin, the halo mass we derive is especially dependent on the bias-to-halo mass relation, due to the steep slope of $M_h(\bgal{})$ for $M_h < 10^{9.5} \msun{}$ as seen in Figure \ref{fig:bias-halomass-curve}.
        Therefore, small differences in the precise modeling of this relation could potentially lead to significant changes in the halo mass; for example, using the theoretical bias-halo mass relation for this bin decreases the inconsistency to $\sim 1.4\sigma$.
        In addition, because our bias-halo mass conversion ignores the stochasticity in the bias-halo mass relation itself, our error estimates for the recovered halo mass are also likely underestimated.

        For the mid- and high-stellar mass bins of $\mathrm{med}(\log_{10} (M_*/M_\odot)) = [9.74, 10.22]$, our \alld{} constraints are consistent with previous observations within a $\sim 1\sigma$ level.
        Due to the shallower slope of the $M_h(\bgal{})$ relation at $M_h < 10^{9.5} \msun{}$, we also expect these bins to be less sensitive to differences in bias-halo mass relation modeling.

        All of our stellar mass bins disagree by $\geq 1 \sigma$ with TNG100-1.
        However, as seen in Figure \ref{fig:hmass-smass-relation} TNG100-1 also disagrees with observational constraints.
        As detailed in \citet{shuntov_2022_cosmos20_shmr}, this observational disagreement is common for most recent simulations above the peak halo mass of $\sim10^{12.5} \msun{}$, and may result from differences in AGN feedback modeling.

        Our \projd{} constraints are consistent with the \alld{} constraints, but with larger errors.
        The weaker constraints from the \projd{} measurement reflect the loss of information from projecting over the line-of-sight dimension.

\section{Conclusion}
\label{sec:conclusion}
In this paper, we carried out a \alld{} cross-correlation analysis between the $2.05<z<2.55$ Lyman-$\alpha$ forest in the sightlines of 320 quasars and galaxies from the CLAMATO survey, and 1642 coeval galaxies from several galaxy redshift surveys within the COSMOS field: CLAMATO itself, zCOSMOS-deep, VUDS, MOSDEF, and 3D-HST.

Using a state-of-the-art estimator, \texttt{Vega}, for the cross power spectrum (Equation~\ref{eq:pk}), we fix the Lyman-$\alpha$ forest parameters and constrain the galaxy linear bias $b_g$, galaxy redshift dispersion/scatter $\sigz$, and systematic redshift offset $\delta_z$ for these heavily-used spectroscopic samples. 

For the surveys that measure redshifts with UV absorption lines (CLAMATO, VUDS, zCOSMOS-deep), we find a systematic redshift offset of $\delta_z \approx +150\,\kms$, indicating that LBG absorption lines are systematically blueshifted relative to the galaxies' true systemic redshift --- consistent with previous work. 
The redshift dispersion $\sigz$ derived for these surveys, on the other hand, are in the range of $\sigz \approx 155 - 274\,\kms$, with a clear dependence on the spectral resolution and signal-to-noise of each survey. 
CLAMATO was conducted with significantly higher resolution than VUDS and zCOSMOS-deep and therefore has the tightest dispersion ($\sigz\approx 155\,\kms$) among the LBG surveys, while VUDS was observed with $>2\times$ longer exposure times than zCOSMOS-deep (on the same telescope and instrument/grating) and therefore exhibits less redshift scatter than the latter.
For the MOSDEF near-IR redshift survey, we find the redshift scatter to be $\sigz \approx 135\,\kms$, which is a greater value than usually expected from redshifts derived from restframe optical nebular emission lines ($\sim 60\,\kms$).
This may be due to finger-of-god redshift distortions, caused by multiple $2 \lesssim z \lesssim 2.5$ COSMOS overdensities which intersect the MOSDEF survey footprint \citep{ata:2022}.
However, we note that this deviation in the MOSDEF $\sigz{}$ is only at the 2.4$\sigma$ level.
Meanwhile, we find a systematic redshift offset of $\delta_z \approx 60\,\kms$, which is roughly $2 \sigma$ from zero, i.e.\ in mild tension with the expectation that nebular emission lines are directly tracing galaxy systemic redshifts. 
However, this is probably due to a statistical fluke from the small number statistics and limited survey volume.
We also find a cross-correlation signal from the 3D-HST survey, but with a large redshift scatter of $\sigz \sim 1400\,\kms$; this is expected given that it is a slitless grism survey with limited spectral resolution.

We then repeat the measurement on a sample of 1150 galaxies with stellar masses estimated from photometric SED fitting in the COSMOS2020 redshift survey.
We split this sample into 3 bins with different stellar masses, and then combine the marginalized posteriors of galaxy bias $b_g$ across the different surveys, for each stellar mass bin. 
For the galaxy samples with median stellar masses $\mathrm{med}(\log_{10} (M_*/M_\odot)) = [9.28, 9.74, 10.22]$, we estimate galaxy linear biases of $b_g=[2.08^{+0.37}_{-0.37}, 3.19^{+0.37}_{-0.37}, 3.78^{+0.36}_{-0.36}]$, respectively.
We repeated the analysis through the \projd{} cross-correlation (i.e.\ projected across the radial or line-of-sight dimension), and found consistent results. 
Our derived galaxy biases are generally consistent with those of \citet{durkalec_2018_vuds_clustering}, who found bias values of $b_g=[1.99\pm0.58, 2.29 \pm 0.64, 2.39 \pm 0.67, 2.84 \pm 0.99] $ in stellar bins of $\log_{10} M_*^{\text{min}} = [8.75, 9.25, 9.75, 10.0]$, respectively. 
The \citet{durkalec_2018_vuds_clustering} result is, to our knowledge, the only other galaxy bias measurement of UV-selected galaxies as a function of their stellar mass at $z\sim 3$, in which the stellar masses are robustly measured through NIR observations. 
Most other published constraints on the bias of LBGs, such as \citet{harikane:2022} have been based on the UV luminosity, making it difficult for us to make one-to-one comparisons. 

Finally, we attempted to place constraints on the underlying halo masses traced by our galaxy samples.
We calibrated our cross-correlation measurements by applying the \texttt{Vega} code to mocks generated from narrow bins of halo masses in the Bolshoi-Planck $N$-body simulations. 
Our results imply that, on the higher stellar mass bins, $\mathrm{med}(\log_{10}{M_*}) =[9.74, 10.22]$ galaxies are associated with typical halo masses of $\mathrm{med}(\log_{10}{M_h}) =[11.65, 12.06]$, respectively. 
These are consistent with the past measurements of \citet{newman-24-latis} and \citet{shuntov_2022_cosmos20_shmr}, as well as the simulation predictions of IlustrisTNG. %
For our lowest stellar mass bin of $\mathrm{med}(\log_{10}{M_*}) = 9.28$, however, we find a corresponding halo mass of $\mathrm{med}(\log_{10}{M_h}) =10.47$. 
This is lower than the observations from COSMOS2020 and the model predictions from IlustrisTNG, at a tension of $2.2$ and $3$-$\sigma$, respectively. This is also inconsistent with the halo mass derived from the LATIS cross-correlation analysis \citep{newman-24-latis} for their lowest stellar mass bin, although they do not go to quite as low stellar masses as we do.
We also note that the LATIS analysis is calibrated using the \texttt{Astrid} cosmological hydrodynamical simulations, and therefore implicitly reflects that particular sub-grid physics model of galaxy formation.
On the other hand, our result from the lowest mass bin is consistent with the corresponding lower mass bins from \citet{durkalec_2018_vuds_clustering}. 
These results imply that galaxy halos with $M_h \sim 10^{10}\,M_\odot$ have experienced significantly stronger star-formation efficiency leading up to Cosmic Noon, than expected from the TNG or UniverseMachine models. 
However, a significant caveat is that our galaxies in the low stellar mass bin are overrepresented by VUDS, i.e. the same data analyzed by \citet{durkalec_2018_vuds_clustering}.
While our \lyaf{} cross-correlation breaks many of the systematic effects and degeneracies inherent in auto-correlation measurements, one can conceive of effects that might affect the estimated bias from both auto- and cross-correlations. For example, if the UV luminosity or star-formation were suppressed in some lower-mass galaxies in overdensities within the VUDS sample, i.e. a similar effect as hinted by \citet{newman2020}, then this would reduce LBG clustering.

There are several ways in which the analysis presented in this paper could be improved upon.
In particular, our stated uncertainties in the halo mass for each stellar mass bin are likely understated due to unmodeled systematics in the bias-halo mass relation.
Due to the degeneracy between $\bgal{}$ and $\sigz{}$, future attempts at combining surveys with heterogenous redshift accuracies/dispersions would benefit from more sophisticated modeling of this relation, perhaps for each survey individually before combining.
Additionally, when propagating our uncertainties on the bias to the halo mass for each bin, we assume that the bias-halo mass relation is completely deterministic.
Since this is not true in practice, this also adds additional uncertainty to our results.
However, we note that our fitted bias and redshift parameters can be directly used as a base for any future sophisticated error modeling analyses.

In our analysis, we have also assumed the \lyaf{} is well-described with the fluctuating Gunn-Peterson approximation (FGPA) of the optical-thin photoionized regime. 
Recent CLAMATO analyses of galaxy protoclusters in COSMOS have suggested that some of the large-scale ($\gtrsim 5\,\mpc$) protocluster gas might be affected by AGN feedback, leading to a departure from FGPA \citep{Kooistra:2022,dong_2023_costco,dong:2024}. 
If this feedback effect turns out to be as strong as suggested by these papers, then future analyses need to incorporate more complex models of the \lyaf.

Our fitted values for the galaxy bias $\bgal{}$ are also inversely proportional to the fixed value of the \lya{} bias \blya{} we choose, as seen from Equation \ref{eq:pk}. 
It is difficult to make a direct comparison between our \blya{} and those found from other analyses, as we use an effective \blya{} from an analysis where HCDs have not been removed. 
However, the \blya{} and $\beta_{\text{\lya{}}}$ parameters found in previous analyses can have non-trivial differences stemming from choices in contaminant modeling; for example, the \blya{} value from SDSS DR14 disagrees by $\sim 20\%$ with the value from SDSS DR12 \citep{blomqvist-2019-ebossdr14-crosscorr}. 
Deviations in \blya{} on this scale may explain the low halo mass in our low stellar-mass bin due to the sensitivity of the \bgal{}-halo mass relation in this bias range. Fortunately, as \blya{} cleanly factors out of the cross-correlation, our \bgal{} values can be easily rescaled for different assumptions of \blya{}.
Differences in the assumed RSD parameter $\beta_{\text{Ly$\alpha$}}$ may also affect our results; however, as this parameter only affects the cross-correlation in orders of $\mu^2 = (\cos{\theta})^2$, differences in $\beta_{\text{Ly$\alpha$}}$ are likely absorbed into the redshift dispersion \sigz{}, and would affect $\bgal{}$ less.

For our galaxy redshift data, we do not restrict ourselves to the highest-quality redshifts, using redshifts with rated confidence levels greater than $\sim 75\%$. 
If many of these medium-confidence redshifts are significantly off, this may dilute the cross-correlation and bias our fits for \bgal{}. 
This would likely primarily impact the low stellar-mass bin in our mass-dependent analysis, and could possibly explain our low halo mass for that bin.
To investigate this possibility, we look at $2 < z < 3$ galaxies within the COSMOS Spectroscopic Redshift Compilation that have been observed by multiple surveys \citep{khostovan-2026-cosmos-compilation}. We find that redshifts with confidence level $\geq 75\%$, when not catastrophically wrong, differ by $\Delta z < 0.01$ from the best redshift on record. At $z = 2.3$, this translates to a radial offset of $< 9$ $\mpc{}$, and so might actually inflate our fitted $\bgal{}$.
While catastrophically-wrong lower-confidence redshifts are still possibly a concern, we believe this is not a major systematic for our dataset.
For example, half of our zCOSMOS-deep dataset consists of $80 \%$ confidence redshifts, but our fitted low-mass $\bgal{}$ is consistent with surveys such as CLAMATO and MOSDEF which are almost entirely high-confidence redshifts.

In addition, while $\bgal{}$ for the low-mass bin of VUDS is an outlier from other surveys, we do not find the distribution of redshift confidence levels for VUDS to be systematically worse than the other surveys.

In the near future, we anticipate carrying out similar analyses with the Galaxy Evolution survey data from the Subaru Prime Focus Spectrograph (PFS; \citealt{greene2022}), which has recently commenced  survey operations at the time of writing.
With a sample size $\sim 40\times$ larger than CLAMATO, we expect precise measurements of the Lyman-$\alpha$ forest-galaxy cross-correlation to fully characterize the stellar mass-halo mass relationship, as well as potentially incorporating other information such as star-formation rates and AGN activity as enabled by the multi-wavelength data available in these fields.

Beyond PFS, upcoming spectroscopic surveys such as DESI-2 will target LBGs at $z\gtrsim 2$ over footprints of thousands of square degrees \citep[e.g.,][]{Ruhlmann-Kleider:2024}, with cross-correlation measurements likely to form a crucial part of the scientific yield from these efforts (see, e.g., \citealt{Herrera-Alcantar:2025} for a pilot project). In combination with stellar masses derived from the Nancy Grace Roman Space Telescope NIR imaging \citep{ObservationsTimeAllocationCommittee:2025} across (hopefully) the same footprint, this will lead to tight constraints on the stellar-halo mass relationship at a critical cosmic epoch.

\section*{Acknowledgements}

    We are indebted to valuable discussions with Martin White.
    We thank the anonymous referee for their valuable comments.
    We used ChatGPT-4.5 to prepare a draft version of the Introduction, which was then carefully fact-checked and edited. We also use ChatGPT for minor typological and grammatical edits elsewhere in the text.
    KGL acknowledges support from JSPS Kakenhi grant Nos. JP18H05868, JP19K14755 and JP24H00241. Kavli IPMU was established by World Premier International Research Center Initiative (WPI), MEXT, Japan. This work was performed in part at the Center for Data-Driven Discovery, Kavli IPMU (WPI), and made use of computing resources at Kavli IPMU. This work was performed in part at Aspen Center for Physics, which is supported by National Science Foundation grant PHY-2210452.
    AFR acknowledges financial support from the Spanish Ministry of Science, Innovation, and Universities (MICIU), under programmes CEX2024001442-S and PID2024-159420NB-C41, as well as support from the European Union (ERC Consolidator Grant, COSMO-LYA, grant agreement 101044612). IFAE is partially funded by the CERCA program of the Generalitat de Catalunya.

\section*{Data Availability}

The Lyman-$\alpha$ forest data used in this analysis was published in \citet{horowitz-22-clamato-dr2} and is publicly available on Zenodo\footnote{\url{https://doi.org/10.5281/zenodo.7524313}}. 
All the galaxy spectroscopic redshifts used in this paper are now, in principle, available in the public compilation of \citet{khostovan-2026-cosmos-compilation}, although we had previously compiled our own sample and cannot guarantee exact reproducibility. 
The code used in this analysis is made available on GitHub\footnote{\url{https://github.com/bzh-bzh/clamato-xcorr}}.

\input{main.bbl}

\appendix

\section{Full mass-dependent fitted galaxy biases}
\label{appendix:mass-split-table}
Here we present the full table of the cross-correlation results with each galaxy survey (Table~\ref{tab:mass-split-params}), split by stellar mass as described in Section~\ref{sec:split-analysis}.
The systemic redshift offset, $\deltaz$, and line-of-sight dispersion, $\sigz$, are fitted globally for all galaxies in  each survey.
The Low-, Mid-, and High-Mass bins have median stellar masses of $\lgmstar = [9.28, 9.74, 10.22]$, respectively.
The error bars on the fitted biases in the table indicate the 16th and 84th percentile limits of the posterior distributions.

        \begin{table*}
            \centering
            \renewcommand{\arraystretch}{1.2}
\begin{tabular}{lll|lll|lll|lll}
                &                                               &                                               & \textbf{}                 & \textbf{Low-mass}                                 &                                                   & \textbf{}                 & \textbf{Mid-mass}                                 &                                                   & \textbf{}                 & \textbf{High-mass}                                &                                                   \\
\textbf{Survey} & \textbf{$\hat{\delta_z}$ [km/s]}              & \textbf{$\hat{\sigma_z}$ [km/s]}              & \textbf{$N_{\text{bin}}$} & \textbf{3D $b_g$}                                 & \textbf{2D $b_g$}                                 & \textbf{$N_{\text{bin}}$} & \textbf{3D $b_g$}                                 & \textbf{2D $b_g$}                                 & \textbf{$N_{\text{bin}}$} & \textbf{3D $b_g$}                                 & \textbf{2D $b_g$}                                 \\ \hline
CLAMATO         & $+150${\raisebox{0.5ex}{\tiny$^{+19}_{-18}$}} & $135${\raisebox{0.5ex}{\tiny$^{+23}_{-22}$}}  & 41                        & $3.30${\raisebox{0.5ex}{\tiny$^{+0.73}_{-0.71}$}} & $2.72${\raisebox{0.5ex}{\tiny$^{+1.03}_{-1.04}$}} & 55                        & $3.38${\raisebox{0.5ex}{\tiny$^{+0.59}_{-0.59}$}} & $3.23${\raisebox{0.5ex}{\tiny$^{+0.91}_{-0.89}$}} & 55                        & $4.73${\raisebox{0.5ex}{\tiny$^{+0.62}_{-0.60}$}} & $3.94${\raisebox{0.5ex}{\tiny$^{+0.90}_{-0.90}$}} \\
MOSDEF          & $+53${\raisebox{0.5ex}{\tiny$^{+24}_{-24}$}}  & $125${\raisebox{0.5ex}{\tiny$^{+31}_{-30}$}}  & 40                        & $2.56${\raisebox{0.5ex}{\tiny$^{+0.80}_{-0.81}$}} & $2.90${\raisebox{0.5ex}{\tiny$^{+1.54}_{-1.47}$}} & 34                        & $1.78${\raisebox{0.5ex}{\tiny$^{+0.93}_{-0.88}$}} & $3.41${\raisebox{0.5ex}{\tiny$^{+1.66}_{-1.62}$}} & 64                        & $3.08${\raisebox{0.5ex}{\tiny$^{+0.70}_{-0.69}$}} & $3.11${\raisebox{0.5ex}{\tiny$^{+1.20}_{-1.21}$}} \\
VUDS            & $+158${\raisebox{0.5ex}{\tiny$^{+47}_{-51}$}} & $191${\raisebox{0.5ex}{\tiny$^{+39}_{-34}$}}  & 151                       & $0.83${\raisebox{0.5ex}{\tiny$^{+0.59}_{-0.49}$}} & $1.21${\raisebox{0.5ex}{\tiny$^{+0.77}_{-0.68}$}} & 103                       & $3.00${\raisebox{0.5ex}{\tiny$^{+0.78}_{-0.74}$}} & $2.96${\raisebox{0.5ex}{\tiny$^{+0.98}_{-1.00}$}} & 70                        & $2.79${\raisebox{0.5ex}{\tiny$^{+0.97}_{-0.92}$}} & $4.15${\raisebox{0.5ex}{\tiny$^{+1.19}_{-1.19}$}} \\
zCOSMOS-deep    & $+201${\raisebox{0.5ex}{\tiny$^{+53}_{-47}$}} & $347${\raisebox{0.5ex}{\tiny$^{+115}_{-81}$}} & 151                       & $2.19${\raisebox{0.5ex}{\tiny$^{+0.91}_{-0.84}$}} & $2.74${\raisebox{0.5ex}{\tiny$^{+1.11}_{-1.09}$}} & 191                       & $4.12${\raisebox{0.5ex}{\tiny$^{+0.86}_{-0.81}$}} & $3.24${\raisebox{0.5ex}{\tiny$^{+0.99}_{-1.00}$}} & 195                       & $3.72${\raisebox{0.5ex}{\tiny$^{+0.77}_{-0.76}$}} & $3.35${\raisebox{0.5ex}{\tiny$^{+0.97}_{-0.99}$}}
\end{tabular}
            \caption{Median values for the velocity parameters fit across all mass bins in the \alld{} mass-dependent analysis, as well as bin galaxy numbers and fitted biases from both \projd{} and \alld{} analyses. We note that because these fits come from fitting a zero-truncated Gaussian to the bias posterior, the lower uncertainty for fits which are truncated is an underestimate of the Gaussian's standard deviation.}
            \label{tab:mass-split-params}
        \end{table*}

\bsp	%
\label{lastpage}
\end{document}